\documentclass{aa}  

\usepackage[utf8]{inputenc}
\usepackage[T1]{fontenc}
\usepackage{multirow}
\usepackage{xcolor}
\usepackage{natbib}
\usepackage{mathtools}
\usepackage{hyperref}
\usepackage{graphicx}
\usepackage{dcolumn}
\usepackage{bm}
\usepackage{siunitx}					
\usepackage{physics}
\usepackage{comment}
\DeclareSIUnit \parsec {pc}
\usepackage{txfonts}
\usepackage[switch]{lineno} 
\usepackage[normalem]{ulem}

\begin{document} 

   \title{Assessing differences between local dust attenuation and point source extinction within the same galactic environments}


\author{
J.~Duarte \inst{1,}\thanks{joao.r.d.duarte@tecnico.ulisboa.pt},
S.~Gonz\'alez-Gait\'an \inst{2,3,1},
A.~Mour\~ao \inst{1},
J.~Rino-Silvestre \inst{1},
M.~Baes \inst{4},
J.~P. Anderson \inst{3,5},
L.~Galbany \inst{6,7},
M.~Stalevski \inst{8,4}
}

\authorrunning{J.~Duarte}
\institute{CENTRA, Instituto Superior T\'ecnico, Universidade de Lisboa, Av. Rovisco Pais 1, 1049-001 Lisboa, Portugal
\and
Instituto de Astrof\'isica e Ci\^encias do Espaço, Faculdade de Ci\^encias, Universidade de Lisboa, Ed. C8, Campo Grande, 1749-016 Lisbon, Portugal
\and
European Southern Observatory, Alonso de C\'ordova 3107, Casilla 19, Santiago, Chile
\and
Sterrenkundig Observatorium, Universiteit Gent, Krijgslaan 281, 9000 Gent, Belgium
\and
Millennium Institute of Astrophysics MAS, Nuncio Monsenor Sotero Sanz 100, Off. 104, Providencia, Santiago, Chile
\and
 Institute of Space Sciences (ICE-CSIC), Universitat Aut\`onoma de Barcelona, Campus UAB, Carrer de Can Magrans, E-08193 Barcelona, Spain
\and
Institut d’Estudis Espacials de Catalunya (IEEC), Mediterranean Technology Park (PMT), Baix Llobregat Campus – UPC, E-08860  Castelldefels (Barcelona), Spain
\and
Astronomical Observatory, Volgina 7, 11060 Belgrade, Serbia
}

   \date{Accepted June 2025/ Received March 2025}
\abstract
{Dust attenuation in galaxies has often been used as a proxy for the extinction of point sources, such as supernovae, even though this approach ignores fundamental differences between the two cases.}
{We present an analysis of the impact of geometric effects and scattering within dusty media on recovered galaxy dust properties.}
{We employed SKIRT, a radiative transfer code, to simulate observations of point sources embedded in dust clouds, as well as spiral and elliptical galaxies. We examined various galaxy morphologies, inclinations, and instrument apertures.}
{We find that in galaxies the scattering of light into the line of sight and the presence of sources at different depths within the galaxy make attenuation fundamentally different from extinction. For a medium with an intrinsic extinction curve slope of $R_V=3.068$, we recover effective attenuation curve slopes, $R_{V_{eff}}$, ranging from $0.5$ to $7$, showing that the two are not analogous, even for local resolved observations. We find that $R_{V_{eff}}$ greatly depends on dust density, galaxy morphology, and inclination, the latter being the most significant. A single simulated galaxy, viewed from different angles, can qualitatively reproduce the well-known relation between attenuation strength, $A_{V_{eff}}$, and $R_{V_{eff}}$ observed for star-forming galaxies. An increase in dust density leads to higher $R_{V_{eff}}$ across all inclinations, which, assuming a correlation between stellar mass and dust density, explains the increase in $R_{V_{eff}}$ with mass observed in star-forming galaxies. We cannot explain the observed differences in $R_{V_{eff}}$ between star-forming (higher $R_{V_{eff}}$) and quiescent (lower $R_{V_{eff}}$) high-mass galaxies, suggesting intrinsically lower $R_V$ values for ellipticals.}
{We conclude that highly attenuated regions of simulated face-on galaxies yield $R_{V_{eff}}$ within a $10\%$ error of the intrinsic extinction $R_{V}$ of the medium, allowing different dust types to be distinguished. For edge-on spirals, the median $R_{V_{eff}}$ for low $A_{V_{eff}}$ regions appears to better approximate the extinction $R_{V}$.}

\maketitle
\section{Introduction}
\label{sec:intro}
Type Ia supernovae (SNe Ia) have long served as fundamental distance indicators for observational cosmology. They have contributed to the measurement of the accelerated expansion of the Universe \citep{Riess98,Perlmutter_99} and to the constraint of various cosmological parameters, such as the Hubble constant, $H_0$, \citep[e.g.,][]{Dhawan_18,Riess_2022,Galbany_2023}. However, when comparing the values of $H_0$ obtained via SNe Ia and other distance indicators of the distance ladder with those obtained from early-time probes, a tension of $\sim5\sigma$ has been reported \citep{planck_2018,Riess_2020}. Although several cosmological model extensions have been proposed as the solution to this tension \citep[e.g.,][]{Karwal_2016,Zhao_2017,Raveri_2020}, the possibility that some systematic error is at least partially responsible has not yet been fully discarded. In fact, recent results show that by using an inverse distance ladder whereby SN distances are calibrated via baryon acoustic oscillation (BAO) data, an $H_0$ compatible with early-time measurements can be recovered \citep{camilleri2024dark}, pointing to the fact that, from the SN Ia point of view, the tension might originate in the standard local distance ladder \citep{wojtak2024}.
    
\par
The process of SN Ia standardization for cosmology relies on a number of empirical corrections based on observed color-luminosity and light-curve shape-luminosity \citep{Phillips_1993,Tripp98} relations, which have usually been assumed to be universal, i.e., the same for all SNe~Ia \citep{Guy_2010}. However, SNe Ia originating in high-mass galaxies have been shown to be more luminous than those originating in low-mass galaxies after the shape and color corrections \citep{Sullivan10,kelly_2010,Lampeitl_2010}. This effect is known as the ``mass step'' and it could hint at the existence of a systematic bias in the standardization, most likely related to the incomplete treatment of the previously mentioned color-luminosity relation, which could potentially influence SN Ia cosmological results.
\par
 On top of its assumption of validity for all SNe~Ia, the color-luminosity relation also combines two different physical effects in the same linear relation: i) intrinsic color-luminosity relations, i.e., fainter SNe Ia are redder; and ii) extrinsic dust effects, i.e., more dust extinction leads to fainter magnitudes and redder SNe through selective absorption. \cite{gaitan_2021} propose the existence of two separate populations of SNe Ia as an explanation for the mass step, with each having different intrinsic color-luminosity properties and dust environment characteristics. \cite{Brout_2021} suggest that host galaxy dust differences are solely responsible for the effect and define two separate $R_V$ populations for low- and high-mass galaxies, finding significantly different values for each one and an overall reduction in the mass step and \cite{Johansson_2021} recover no mass step when using an individual best-fit $R_V$ for each standardized SN Ia. Similarly, \cite{wiseman_2022} find that using separate $R_V$ populations for younger and older galaxies results in a similar improvement, while \cite{popovic2024modelling} conclude that the mass step can, in part, be attributed to luminosity differences related to the age of the SN Ia progenitor and to its local environment. \cite{Duarte_2022} find that SN host galaxy properties can be used to define a ``dust step,'' which nevertheless fails to fully account for the mass step. It has also been shown that apparent galactic dust properties can vary considerably between different galaxies and that they strongly correlate with environmental properties such as the stellar mass and the stellar age of the population in question \citep[e.g.,][]{Garn_2010,Zahid_2013,Salim_2018,Duarte_2022}. As such, assuming a standard universal dust effect for SN Ia might prove inadequate in correcting the color-luminosity relation and could ultimately be at least partially responsible for the observed mass step.
\par
In an effort to improve SN Ia standardization, many methodologies have recently been proposed \citep[e.g.,][]{Brout_2021,wiseman_2022,Duarte_2022}, relying on the assumption that galactic dust properties can be used as an adequate proxy for SN Ia extinction. These approaches, while in some cases presenting a degree of success in reducing the Hubble residuals for the calibrated SN, might ultimately prove to be incomplete, given that attenuation and extinction are not the same. While they appear similar, the reddening laws describing the effects of dust on SN and other point source observations, also known as extinction laws, are fundamentally different from the reddening laws affecting integrated light observations such as galaxies, also known as attenuation laws \citep{Krugel_2009,Salim_2020}. This also means that the parameters usually used to describe dust effects, while sharing the same nomenclature in most of the literature, must be interpreted with care and in different ways.
\par
As light travels through the interstellar medium (ISM) and interacts with dust particles, it experiences absorption and scattering effects, which are especially prevalent for blue wavelengths \citep{Calzetti_1994}. These effects lead both to a dimming and a reddening of the original spectral energy distribution, which can be parameterized by an empirical reddening law \citep[e.g.,][]{Cardelli_1989,Fitzpatrick_1999,Calzetti_2001}.

\par
For a point source, the total extinction for the $V$ band, $A_V$, which serves as a normalization for the extinction curve, describes the dust column density in the optical path of the observed light. Likewise, the color excess for the $B-V$ color index, $E(B-V)$, refers only to the reddening experienced along that single line of sight. Therefore, the total-to-selective extinction ratio, $R_V=A_V/E(B-V)$, which describes the slope of the extinction curve, is only a function of the optical properties of the dust grains themselves \citep{Draine_2011}.
\par
For an extended stellar population or a galaxy, this picture is complicated as a result of two primary factors. On the one hand, the necessity of using a larger aperture for galaxy observations makes the light scattered back into the line of sight a relevant component of the overall observed flux, with the strength of this scattered component being dependent on the morphology and inclination of the galaxy, as was demonstrated by \cite{Chevallard_2013}. Those authors show that galaxy inclination can drastically impact attenuation properties, meaning that identical galaxies viewed from different angles can produce wildly different attenuation curve strengths and slopes. On the other hand, the interspersion of stars and dust lanes leads to varying column densities inside the galaxy, along with different stellar light contributions. \cite{Viaene_2017} found that the thicker the dust lane, the more the emission is dominated by stars located in the outermost regions of the galaxy, which are not meaningfully attenuated. 
\par
When taken together, the above effects mean that, in the case of a galaxy, the values measured for $A_{V_{eff}}$ and $R_{V_{eff}}$ stop being representative of the true optical properties of the medium along the line of sight and should instead be understood as effective values. As such, they cannot be easily related to the extinction experienced by SN light, even if the local dust environment is properly mapped \citep{Duarte_2022}.

\par

Although this has long been known, it has only recently permeated the SN field. \cite{Chevallard_2013} and \cite{Viaene_2017} show that a deeper knowledge of the underlying galaxy geometry is needed to properly describe dust attenuation. \cite{Duarte_2022} show that the SN Ia standardization returns much higher Hubble residuals when galactic dust properties are used to directly constrain the color luminosity correction. \cite{popovic2024modelling} conclude that dust in a SN line of sight cannot be accurately described using its host galaxy dust properties. 
\par

\par
In this work we present a comprehensive analysis of how point source extinction relates to galactic attenuation. We rely on radiative transfer simulations to systematically study how factors such as instrument aperture and galaxy inclination, morphology and dust composition influence the recovered values of $A_{V_{eff}}$ and $R_{V_{eff}}$ and how these relate to the real optical properties of the medium. More specifically, we seek to understand under which conditions, if any, a host galaxy dust measurement can be said to truly reflect the extinction experienced by a local point source, such as a SN, and therefore be used as its proxy. We use the radiative transfer code SKIRT \citep{Camps_2015,CAMPS_2020} to simulate observations for both point sources and a variety of galaxies (spiral and elliptical), observed from multiple angles and with differing apertures, the details of which are presented in Section \ref{sec:methods}. In Section \ref{sec:results}, we present and discuss our simulation results and compare them with observational data, focusing on how well purely geometrical effects can explain observed $A_{V_{eff}}-R_{V_{eff}}$ relations. In Section \ref{sec:analysis}, we analyze the link between dust properties and stellar mass, an environmental property related to SN Ia standardization, and to what extent it can be explained by our simulations. We also discuss how dust attenuation data could be used to quantitatively recover true extinction $R_V$ values, as well as the impacts of the source spectrum on the recovered dust properties. Finally, in Section \ref{sec:conclusions}, we summarize our main conclusions.

\section{Methods}
\label{sec:methods}
In this work, we made use of SKIRT, a powerful radiative transfer code that uses the Monte Carlo method \citep{Camps_2015,CAMPS_2020}\footnote{\url{https://skirt.ugent.be/root/_home.html}}. This code relies on the simulation of the path of a number of photon packets emitted from a light source through a gridded astrophysical dusty medium. By taking into account both absorption and scattering effects, the code is able to stochastically simulate a view of the model from any orientation, as observed by a particular instrument. While it is also possible to simulate the contributions of dust emission, we did not do so, as it is not particularly relevant for the wavelengths in question in this work.

\par
 We are interested in simulating the basic geometries of both galaxies and point sources. With this goal, we focus on three highly abstracted models: 1) a point source immersed within a galactic dust cloud; 2) a spiral galaxy-like model; 3) an elliptical galaxy-like model. The point source system is mainly used as an ideal benchmark for comparison with the galaxy models.
\par
The simulated astrophysical structure in a SKIRT model is made up of two systems, a light source and a dust medium system. The spectral energy distribution (SED) for the source system was defined based on the simple stellar populations (SSPs) described by \cite{Bruzual_2003}, assuming a \cite{chabrier_2003} initial mass function (IMF), with a metallicity $Z=0.02$ and an age of $t_{age}=5$ Gyr.\footnote{Given that SKIRT simulations output both the simulated observed flux and the total emitted flux, we can directly compute $A_{V_{eff}}$ and $R_{V_{eff}}$, meaning that the recovered dust properties are free from any degeneracies and that they are not impacted by the SED chosen for the source system.}. To simulate aperture photometry observations, $10^8$ photon packets were used, while to simulate full-galaxy field observations (with a pixel size corresponding to a $25$ pc resolution), $10^{11}$ photon packets were used instead\footnote{The value of $10^{11}$ photon packets, despite being quite high, was chosen because, for any lower number, severe stochastic fluctuations in $R_{V_{eff}}$ were present for the majority of the pixels.}. We made use of SKIRT's ``ExtinctionOnly'' mode given that, as was stated above, we are not interested in dust emission.
\par
By default, a \cite{Zubko_2004} dust mix model was used for the dust medium, composed of bare silicate, graphite, and polycyclic aromatic hydrocarbon (PAH) dust grains, with half of the PAH grains being neutral and half being ionized. A Mathis-Rumpl-Nordsiek (MRN) dust mix, composed of bare silicate and graphite \citep{Mathis_1977}, was also used for a specific example, discussed in Section \ref{sec:analysis_local}. The simulation outputs both the emitted and the observed flux, meaning that we can obtain accurate values for both $A_V$ and $R_V$ ($A_{V_{eff}}$ and $R_{V_{eff}}$, if we are in an attenuation scenario). While these two parameters are not ideal for a complete analysis of extinction and attenuation curves, given that they are restricted to the behavior of only two wavelengths, they nevertheless remain one of the most common parametrizations for dust effects, both in SN and galaxies.

\par
We simulated observations from a variety of angles and apertures. We simulated global aperture photometry for an array of aperture SED detectors located at a distance of $10$ Mpc from the center of the model. The detectors were positioned at different angular positions, ranging from a face-on view of the galaxy at $0^\circ$ (along the z axis) to an edge-on view at $90^\circ$ (along the R plane). The detector aperture radius was varied to examine different observing conditions. We also simulated full-galaxy field data for a detector that records the SED for each pixel in the field of view, also located at a distance of $10$ Mpc from the center of the model. A physical pixel size corresponding to a $25$ pc resolution was used.
    
\subsection{Point source in a dust cloud model}

\label{sec:point_sources}
The point source system was defined as a point source located in the center of a galactic dust cloud, normalized in such a way that its total luminosity in the wavelength range of $0.1\si{\micro\meter}<\lambda<1\si{\micro\meter}$ was $L=10^9 L_{\odot}$, similar to the energy output of a SN around peak.
\par

Based on \cite{Baes_2000}, the dust cloud system density profile was described by a flattened spheroidal geometry given by Eq. \ref{eq:sphere}:

\begin{equation}
    \rho(R,z)=\frac{1}{q} \ \rho \left(\sqrt{R^2 + \frac{z^2}{q^2}}\right),
    \label{eq:sphere}
\end{equation}

\noindent
where $q=0.5$ is the flattening parameter and $\rho$ is the \cite{Plummer_1911} density profile, given by Eq. \ref{eq:plummer}:

\begin{equation}
    \rho(r)=\rho_0 \left(1 + \frac{r^2}{c^2} \right)^{-5/2},
    \label{eq:plummer}
\end{equation}
\noindent
where $c=4000$ pc is the scale length. The normalization, $\rho_0$, was varied to produce different optical depths in the dust system.

\subsection{Spiral galaxy model}
\label{sec:spirals}
The light source system for the spiral galaxy was divided into two components: a disk and a bulge. The disk was described by a disk geometry characterized by a double-exponential profile given by Eq. \ref{eq:exp_disk}:

\begin{equation}
    \rho(R,z)=\rho_0\exp{-\frac{R}{h_R}-\frac{\abs{z}}{h_z}},
    \label{eq:exp_disk}
\end{equation}
\noindent
where $h_R=4000$ pc is the scale length and $h_z=350$ pc is the scale height. The normalization, $\rho_0$, was defined such that the total luminosity of the disk in the wavelength range of $0.1\si{\micro\meter}<\lambda<1\si{\micro\meter}$ was $L=5\times10^{9}L_{\odot}$. These values roughly correspond to the ones obtained by \cite{De_Geyter_2014} for a sample of observed spiral galaxies. As we were interested in keeping the model as simple as possible, we did not simulate spiral arms.

\par
The bulge component was described by a flattened spheroidal geometry, defined by Eq. \ref{eq:sphere} with a flattening parameter of $q=0.7$ and a \cite{sersic_63} $\rho(r)$ density profile given by Eq. \ref{eq:sersic}:

\begin{equation}
    \rho(r)=\rho_0 \ S_n\left(\frac{r}{r_{eff}}\right),
    \label{eq:sersic}
\end{equation}
\noindent
 where $r_{eff}=1600$ pc is the effective radius, $S_n(s)$ is the Sérsic function of order $n$ and $n=2$ is the Sérsic index \citep{sersic_63}. Once again, these values roughly correspond to those obtained by \cite{De_Geyter_2014}. The normalization, $\rho_0$, was defined such that the total luminosity of the bulge in the wavelength range of $0.1\si{\micro\meter}<\lambda<1\si{\micro\meter}$ was $L=3\times10^{9}L_{\odot}$.

The dust medium geometry was taken to be the same as that of the spiral source disk, described by a disk characterized by a double-exponential profile as defined by Eq. \ref{eq:exp_disk}, with the normalization, $\rho_0$, being varied to produce simulated galaxies with different optical depths. We used a scale length parameter $h_R=4000$ pc and a scale height parameter $h_z=350$ pc. Even though the dust disk is typically found to be thinner than the stellar disk \citep[e.g.,][]{De_Geyter_2014}, we opted to use disks with the same thickness. We did so because we were primarily interested in the interaction between light and dust and thus wanted to limit the number of stars effectively outside the dust disk. However, we find that using a thinner dust disk does not significantly alter the results discussed below.

\subsection{Elliptical galaxy model}
\label{sec:ellipticals}

The light source system for the elliptical galaxy was described by a flattened spheroidal geometry, defined by Eq. \ref{eq:sphere} with a flattening parameter of $q=0.5$, and a \cite{sersic_63} $\rho(r)$ density profile, as described by Eq. \ref{eq:sersic} with effective radius $r_{eff}=4000$ pc and a Sérsic index $n=4$. The values were selected to roughly follow those cited by \cite{Beifiori_2012} for a sample of observed elliptical galaxies. The normalization, $\rho_0$, was defined such that the total luminosity in the wavelength range of $0.1\si{\micro\meter}<\lambda<1\si{\micro\meter}$ was $L=10^{10}L_{\odot}$.
\par
Once more following \cite{Baes_2000}, the dust system for the elliptical galaxy was described by a flattened spheroidal geometry, defined by Eq. \ref{eq:sphere} with a flattening parameter of $q=0.5$ and a \cite{Plummer_1911} $\rho(r)$ density profile, as described by Eq. \ref{eq:plummer} with $c=4000$ pc. The normalization, $\rho_0$, was varied to produce different optical depths in the dust system.

\section{Results}
\label{sec:results}
\subsection{Point source model}

\label{sec:analysis_point_source}
In this section we present an analysis of simulated photometric observations for the point source model. In particular, we analyze the impact of different apertures on the dust properties recovered for the model and the transition between extinction and attenuation regimes.
\par
\begin{figure}
	\centering
	\includegraphics[width=0.5\textwidth]{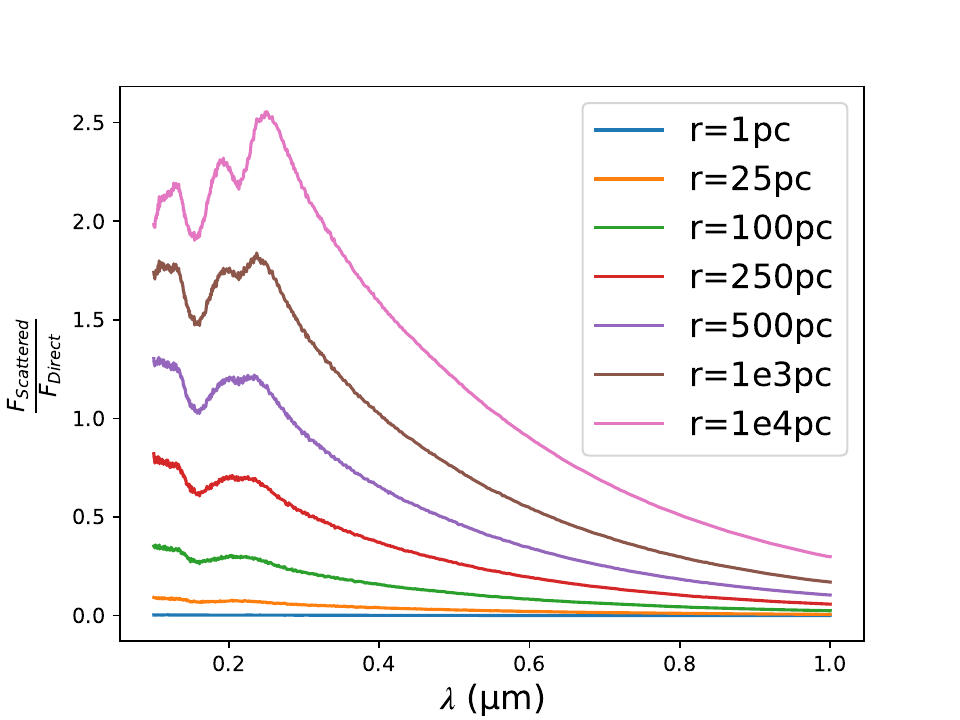}
	\caption{Ratio between the scattered (into the line of sight) and direct fluxes as a function of the wavelength $\lambda$ for face-on observations of a simulated point source, immersed in a galactic dust cloud. Simulated observations for a variety of aperture radii are shown. The cloud dust density has been normalized so that the total optical depth along the R axis is $\tau_{V_R}=5$.}
\label{fig:scattered_fluxes}
\end{figure}

The observed flux can be divided into two components: 1) the direct flux, comprising photon packets that reach the detector without interacting with dust; 2) the scattered flux, comprising packets that have gone through at least one scattering event, ending up in the line of sight. Fig. \ref{fig:scattered_fluxes} shows the ratio between these two fluxes for face-on observations of a point source inside a galactic dust cloud as seen by simulated instruments with different aperture radii. The dust density profile was normalized so that the total optical depth along the total R axis is $\tau_{V_R}=5$.

\par
 As expected, as the size of the aperture used for the observations increases, so does the relative contribution of the scattered flux to the total flux. If the aperture is small enough, as in the case with a radius of $1$ pc, the scattered component is virtually non-existent and we are left with only the direct flux component. Clearly, the larger apertures place us in the attenuation regime, meaning that the recovered dust properties should be understood as effective, as they are no longer representative of the actual optical properties of the medium along the line of sight.
 \par
\begin{figure}
	\centering
	\includegraphics[width=0.5\textwidth]{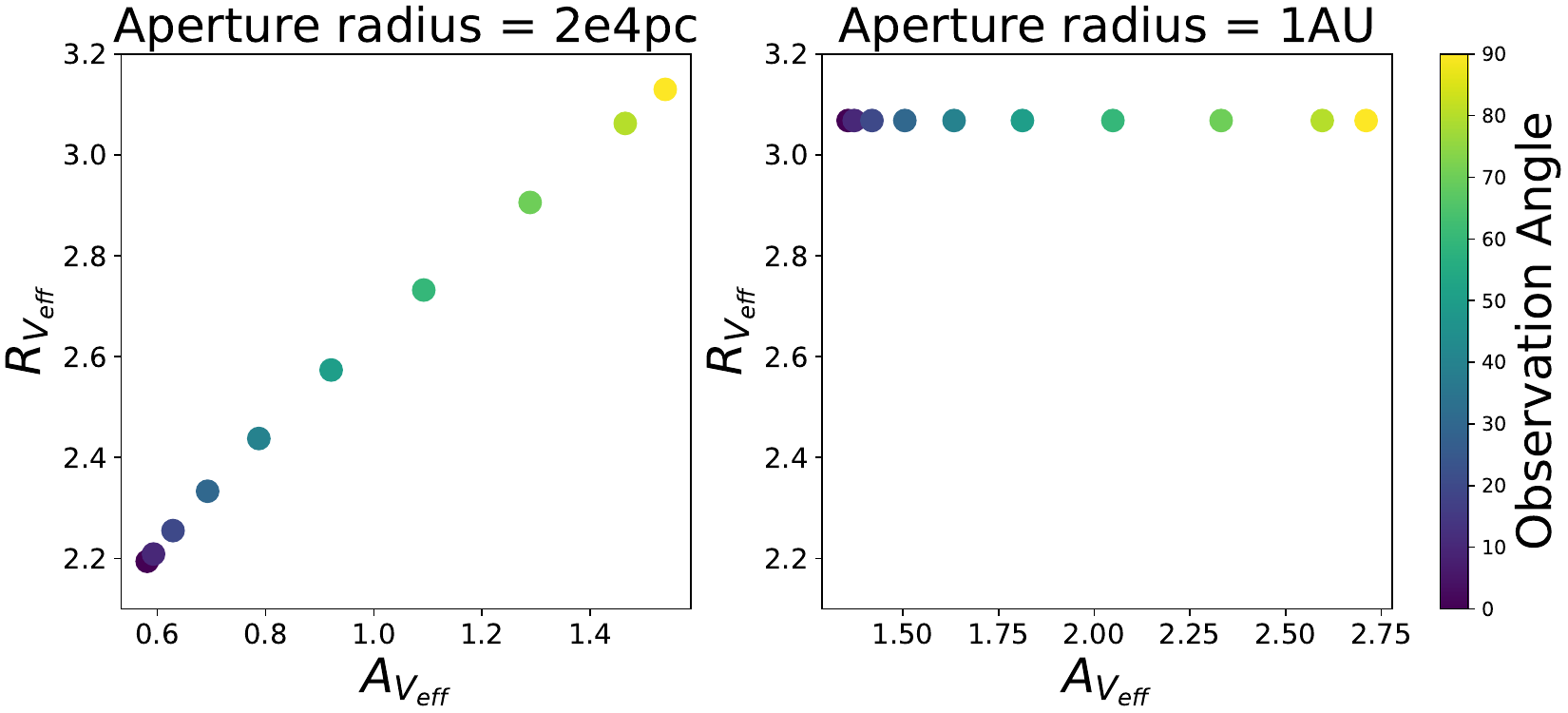}
	\caption{Values of effective $R_{V_{eff}}$ as a function of $A_{V_{eff}}$ for a simulated point source immersed in a galactic dust cloud. Different colors denote different observation angles, from $0^\circ$ (face-on) to $90^\circ$ (edge-on). The results from two different instruments are presented: One with an aperture radius of $1$ AU (left) and another with an aperture radius of $20$ kpc (right). The cloud dust density has been normalized so that the total optical depth along the R axis is $\tau_{V_R}=5$.}

\label{fig:point_source}
\end{figure}

 This is further illustrated in Fig. \ref{fig:point_source}, which shows, as an example, the values of $A_{V_{eff}}$ and $R_{V_{eff}}$ for a variety of observation angles when using two different simulated instruments, with aperture radii of $1$ AU (left) and $20$ kpc (right), respectively. As hinted before, the observed behavior greatly depends on the aperture used. For the smaller aperture, we recover a constant value of $R_V=3.068$ that does not depend on the observation angle and matches exactly the value of the extinction curve slope for the Zubko dust mix $R_{V_{Zubko}}$, as directly computed from the material extinction cross sections outputted by SKIRT. Likewise, the values of $A_V$ can be taken as true indicators of dust extinction, as they correctly reflect the amount of light that has been absorbed or scattered away from the line of sight.
 \par
 For the larger aperture, we note that the recovered $A_{V_{eff}}$ values are $57\%$ to $43\%$ lower than the respective true extinction $A_V$. This is a result of the scattered light that is now contributing to the observations, as is seen in Fig. \ref{fig:scattered_fluxes}. Moreover, we note that the relation between $A_V$ and $A_{V_{eff}}$ is close to linear, indicating that within the same dust cloud the scattered and absorbed fluxes scale consistently at this wavelength. Additionally, we find that $R_{V_{eff}}$ is highly dependent on the galaxy inclination, ranging from $R_{V_{eff}}=2.212$, for face-on observations, to $R_{V_{eff}}=3.120$, for edge-on observations. We conclude that, for this particular example, an increase in the observation aperture can result in a $28\%$ deviation from the intrinsic optical extinction $R_V$.

\begin{figure}
	\centering
	\includegraphics[width=0.5\textwidth]{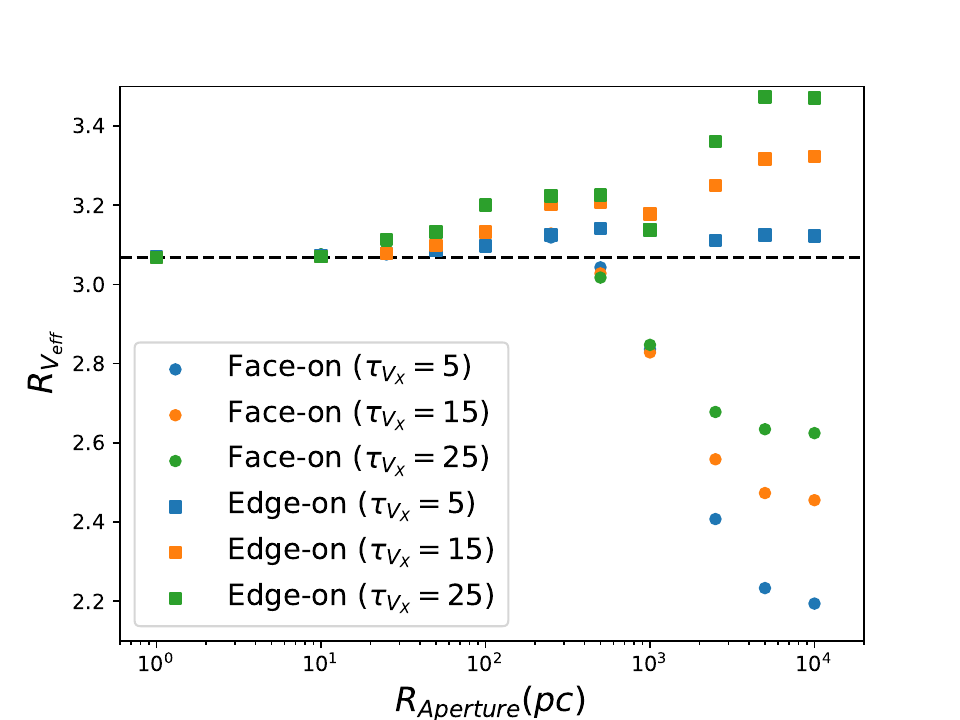}
	\caption{Values of effective $R_{V_{eff}}$ as a function of aperture radius for a simulated point source immersed in a galactic dust cloud. Results are shown in blue for $\tau_{V_R}=5$, orange for $\tau_{V_R}=15$ and green for $\tau_{V_R}=25$. Face-on results are represented by dots, while edge-on results are represented by squares. The dashed black line reflects the true $R_V$ of the medium.}
\label{fig:rv_aperture}
\end{figure}

We can refine our analysis by looking at the $R_{V_{eff}}$ values recovered for each of the apertures discussed in Fig. \ref{fig:scattered_fluxes}, which are plotted in Fig. \ref{fig:rv_aperture}. In addition to the $\tau_{V_R}=5$ normalization discussed above, results for two other dust densities, $\tau_{V_R}=15$ and $\tau_{V_R}=25$, are presented, for both face-on and edge-on observations. We find that, for the current setup, any aperture with a radius larger than $25-50$ pc results in a $R_{V_{eff}}$ that deviates from $R_{V_{Zubko}}$. In particular, as was stated above, we confirm that galaxy inclination has a great impact on both $A_{V_{eff}}$ and $R_{V_{eff}}$, as face-on and edge-on simulated observations differ greatly from each other. 
\par
\begin{figure}
	\centering
	\includegraphics[width=0.5\textwidth]{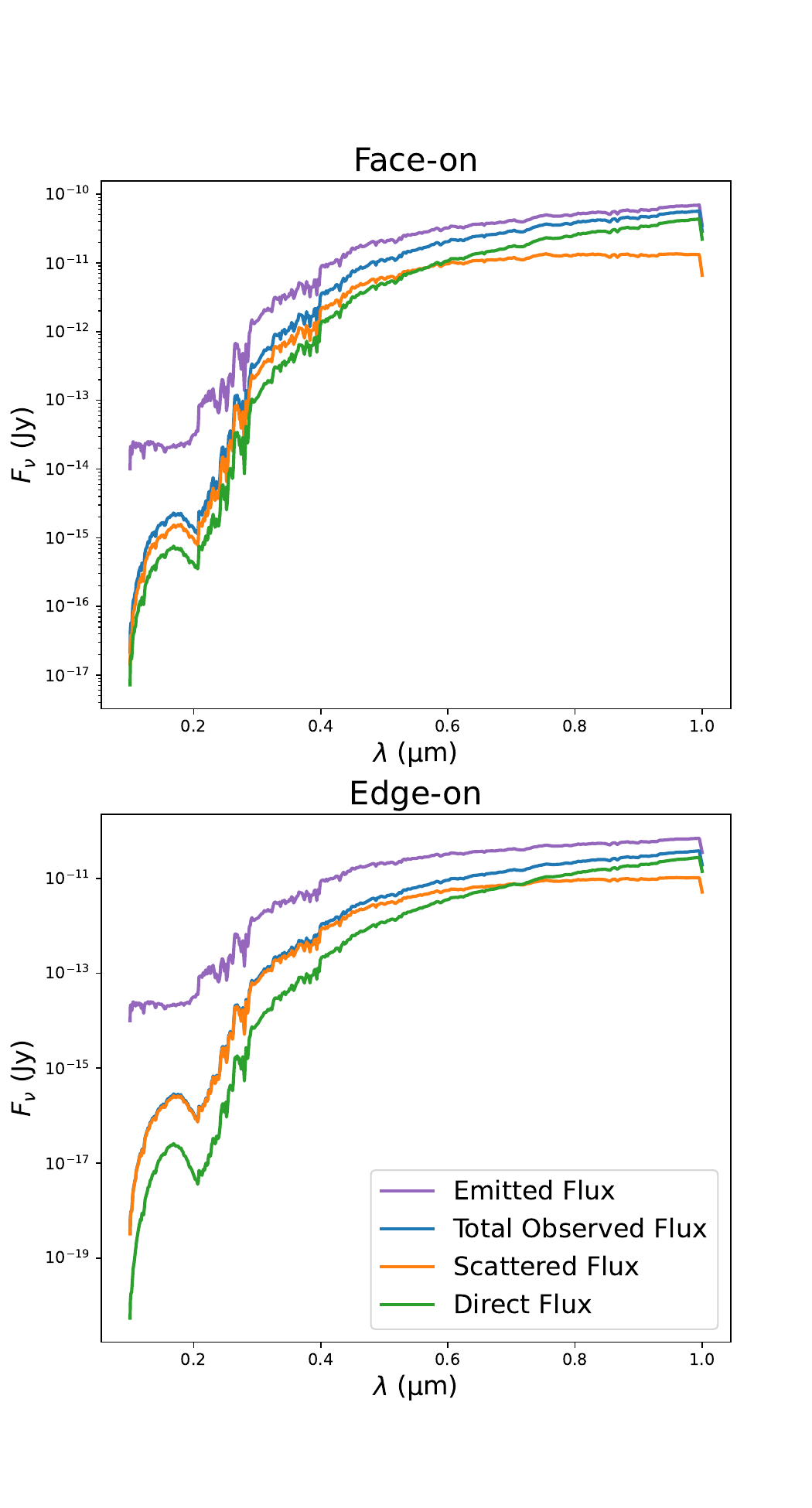}
	\caption{Flux components for face-on (top) and edge-on (bottom) simulated observations of a point source immersed in a galactic dust cloud with $\tau_{V_R}=5$. The observations were simulated for an aperture radius of $r=20$ kpc. The emitted, total observed, scattered, and direct fluxes are shown in purple, blue, orange, and green, respectively.}
\label{fig:fluxes}
\end{figure}

The differences in the simulated $R_{V_{eff}}$ arise from the interplay between the direct and scattered fluxes. In general, the scattered flux tends to dominate the bluer wavelengths, with the direct flux dominating the redder wavelengths, as is shown in Fig. \ref{fig:fluxes}. As we move from a face-on view to an edge-on view, the scattered flux becomes even bluer when compared to the direct flux. Overall, this results in a flatter edge-on attenuation curve, boasting higher values of $R_{V_{eff}}$ than those observed for the face-on view, as is seen in Fig. \ref{fig:rv_aperture}. This effect is primarily driven by the thickening of the optical light path for the edge-on observations, which increases the interaction probability between light and the medium, contributing to the absorption and, most relevantly, the scattering of more blue light away from the direct flux. Although this increase in optical depth is also expected to make the scattered flux overall redder, this component still becomes bluer when compared to the direct flux. Likewise, if the dust density is increased, we expect an increase in $R_{V_{eff}}$ across all observation angles, which is precisely what is observed from the simulations. As such, a higher dust column density actually brings the face-on $R_{V_{eff}}$ values closer to $R_{V_{Zubko}}$, while it increases the deviation for the edge-on values.

\par
 We highlight that the results discussed in this section are independent of the SED chosen for the light-source system and thus reflect the composition and geometry of the dust medium. In fact, repeating the simulations for a light source with
$t_{age}=10$ Myr, we recover the same exact values of $A_{V_{eff}}$ and $R_{V_{eff}}$, apart from expected stochastic variations inherent to the method.
\par
We also note that our point source model is not representative of SN observations. While the angular resolution of most ground based SN Ia surveys would place them above our threshold for the extinction regime, the fact is that, due to their short lifetimes, light can only effectively travel and scatter within a small vicinity around the transient (roughly $0.1$ pc in the first four months after explosion). This means that, unless surrounded by a very dense and exotic dust cloud that could give rise to light echoes, as was described in \cite{Wang_2008} and \cite{Bulla_2018}, the SN-dust interaction should always be properly described by extinction. These SN light echoes are usually only visible at very late stages in the SN evolution. However, the scattering effect described above might be important for other long-lived point sources like stars (e.g., Cepheids) or even compact stellar clusters, taking into account that their source luminosity is fainter than a SN.

\subsection{Galaxy models}
Having established a baseline with point source observations, in this section we present an analysis of simulated photometric observations for galaxy models. This differs from the previous example in that an extended light-source distribution is used instead of a single point source. It is particularly useful to look at two different observation scenarios: global observations, resulting from aperture photometry on the entire galaxy, and local observations, resulting from a full-galaxy pixel-to-pixel 2D map.

\subsubsection{Global observations}
\label{sec:aperture_galaxy}
\begin{figure*}
	\centering
	\includegraphics[width=\textwidth]{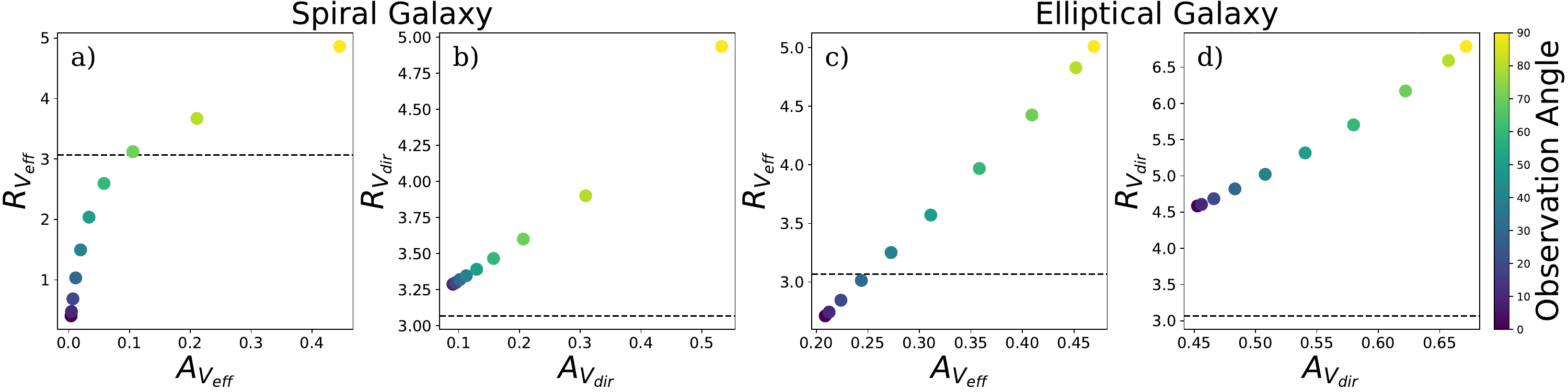}
	\caption{Values of $R_{V_{eff}}$ as a function of $A_{V_{eff}}$ (panels a and c) and the corresponding values of $A_{V_{dir}}$ and $R_{V_{dir}}$ (panels b and d) for a simulated spiral galaxy (left plots) and a simulated elliptical galaxy (right plots). $A_{V_{dir}}$ and $R_{V_{dir}}$ were calculated from only the direct flux component, ignoring the scattered flux component. Different colors denote different observation angles, from $0^\circ$ (face-on) to $90^\circ$ (edge-on). An aperture radius of $20$ kpc was used. The galaxy dust density has been normalized so that the total optical depth along the R axis is $\tau_{V_R}=5$. The dashed black line reflects the true extinction $R_{V_{Zubko}}$.}
    
\label{fig:av_vs_rv_norm5}
\end{figure*}

Fig. \ref{fig:av_vs_rv_norm5} shows the simulated values of $R_{V_{eff}}$ as a function of $A_{V_{eff}}$ for both a spiral and an elliptical-like galaxy model, observed from a variety of angles with an aperture radius of $20$ kpc. In both models, the dust density profile was normalized so that the total optical depth along the R axis is $\tau_{V_R}=5$. 

\par
Generally, for both galaxy types, $R_{V_{eff}}$ follows a trend similar to that observed for the largest aperture in Fig. \ref{fig:point_source}. However, the two morphologies of the galaxies result in noticeable differences, both in terms of the shape of the $A_{V_{eff}}$-$R_{V_{eff}}$ trend and the $R_{V_{eff}}$ value ranges. In particular, for the spiral galaxy, we find that values can range from $R_{V_{eff}}=0.41$ for face-on observations\footnote{The points with $R_{V_{eff}}<1$ are most likely unphysical simulation artifacts, stemming from extremely low values of $A_{V_{eff}}$.}, to $R_{V_{eff}}=4.86$ for edge-on observations. Likewise, we find that for the elliptical galaxy values range from $R_{V_{eff}}=2.71$, for face-on observations, to $R_{V_{eff}}=5.01$, for edge-on observations. In addition, the observation angle for which $R_{V_{Zubko}}$ is recovered varies greatly between the two, being $\sim70^{\circ}$ for the spiral and $\sim30^{\circ}$ for the elliptical. 

\par
As was mentioned in Section \ref{sec:analysis_point_source}, the trend between $A_{V_{eff}}$ and $R_{V_{eff}}$ is primarily driven by the scattering of light into the line of sight. This scattered component, when compared to the direct flux, is redder for the face-on observations, lowering the observed $R_{V_{eff}}$, and bluer for the edge-on view, resulting in higher $R_{V_{eff}}$. In addition to this, the fact that the sources now extend to different depths and contribute to the same optical path means that $R_{V_{eff}}$ deviates even more from $R_{V_{Zubko}}$. 
\par
We can better understand this light source distribution effect by looking at the attenuation properties that we would recover if only contributions from the direct flux were taken into account. As such, for the same two galaxies analyzed in Fig. \ref{fig:av_vs_rv_norm5}, we plot the values of $A_{V_{eff}}$ and $R_{V_{eff}}$ that one obtains if the scattered flux is ignored and only the direct flux is taken into account, which we denote as $A_{V_{dir}}$ and $R_{V_{dir}}$, respectively. These results are shown in Fig. \ref{fig:av_vs_rv_norm5}, from which it becomes clear that, without the reddening effect of the scattered component to counterbalance its effects, the direct flux actually results in higher values of $R_{V_{dir}}$. 

\par

From the above, we conclude that the reddening caused by changes in the direct optical paths behaves similarly to the reddening caused by excess scattered light, with $R_{V_{eff}}$ increasing with the observation angle. In this case, this increase is due to the fact that, as we move from a face-on to an edge-on orientation, there is an increase in the number of different optical paths in our observation, as we are probing much deeper into the dust cloud. This also explains why the spiral galaxy face-on direct flux observations in Fig. \ref{fig:av_vs_rv_norm5} produce the closest results to $R_{V_{Zubko}}$, as this is the scenario in which the sources are most concentrated, given the narrower depth of the galaxy. We also find that, as is observed for the point source, the $A_{V_{eff}}$ values obtained for only the direct flux are systematically higher than those obtained from the full observed flux, as expected.

\par

\begin{figure*}
	\centering
	\includegraphics[width=\textwidth]{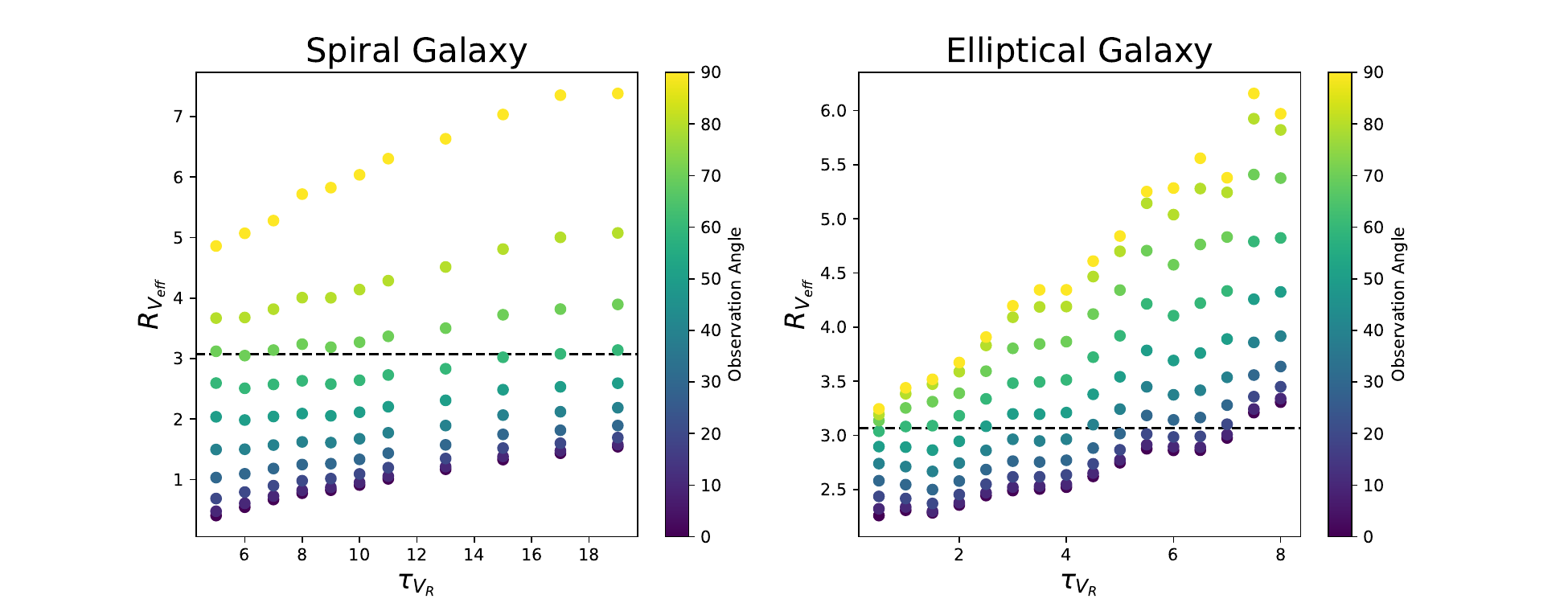}
	\caption{Values of $R_{V_{eff}}$ as a function of the total optical depth normalization along the R axis $\tau_{V_R}$, for a spiral (left) and elliptical galaxy (right). Different colors denote different observation angles, from $0^\circ$ (face-on) to $90^\circ$ (edge-on). An aperture radius of $20$ kpc was used. The dashed black line reflects the true extinction $R_{V_{Zubko}}$.}
\label{fig:av_vs_rv_norms}
\end{figure*}

The effects of considering different dust densities on the effective dust properties are presented in Fig. \ref{fig:av_vs_rv_norms}. Similarly to the results shown in Fig. \ref{fig:rv_aperture}, we see that the amount of dust in a galaxy has a great impact on the attenuation properties, namely by increasing the values of $R_{V_{eff}}$ across all observation angles, similarly to what was discussed in Section \ref{sec:analysis_point_source}. For the normalizations considered, we find that, for spiral galaxies, the values of $R_{V_{eff}}$ range from $0.41$ to $1.54$, at their lowest, and from $4.86$ to $7.38$ at their highest. Likewise, for elliptical galaxies, the values of $R_{V_{eff}}$ range from $2.26$ to $3.31$, at their lowest, and from $3.24$ to $6.16$ at their highest.

\par
This dust density effect is, however, still secondary to the variation in the observation angle, which produces a much larger change in $R_{V_{eff}}$. As such, the effective dust properties of a galaxy appear to be more a function of geometry and morphology than of the dust abundance itself. In practice, this means that, when using global galaxy aperture photometry, the true optical properties of the dust medium, namely $R_{V_{Zubko}}$, become obfuscated and cannot be easily recovered.

\begin{figure}
	\centering
	\includegraphics[width=0.5\textwidth]{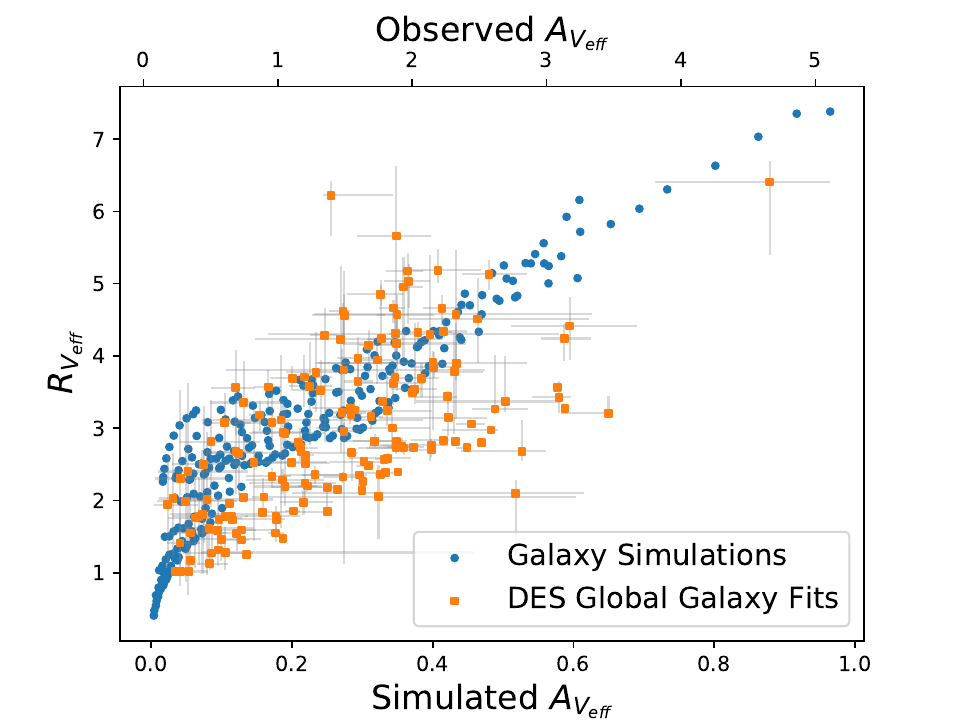}
	\caption{Values of $R_{V_{eff}}$ as a function of $A_{V_{eff}}$ for the simulated galaxies presented in Fig. \ref{fig:av_vs_rv_norms} (blue circles, lower axis) and DES global galaxy fits from \cite{Duarte_2022} (orange squares, upper axis). The plot illustrates the same general trend, even though the $A_{V_{eff}}$ values are not compatible.}
\label{fig:des_comp}
\end{figure}

In Fig. \ref{fig:des_comp}, we present a comparison between the $A_{V_{eff}}$-$R_{V_{eff}}$ trend obtained from the various simulated observations detailed in Fig. \ref{fig:av_vs_rv_norms} and the trend obtained for Dark Energy Survey \citep[DES;][]{Abbott_2018} global galaxy fits, as discussed in \cite{Duarte_2022}. The plot covers different galaxy types, dust densities, and orientations. We find that, with the exception of the absolute $A_{V_{eff}}$ values, our simulations are able to fully reproduce the observed behavior. The mismatch in $A_V$ might actually stem from a bias in the DES $A_{V_{eff}}$ values of \cite{Duarte_2022}, which were obtained using the Prospector Python package \citep{Johnson_2021}. These $A_{V_{eff}}$ values are found to be consistently higher than those recovered for similar galaxies by, for example, \cite{Zahid_2013} and \cite{Salim_2018}, which are much closer to the values obtained from our simulations. We emphasize, however, that the relation between the two quantities is very well explained with our simulations. The increased level of scatter exhibited by the real data is most likely due to the fact that a real sample is expected to reflect a wide mixture of different morphologies and dust types.
\par
Our results are also consistent with simulated observations of galaxies with different inclinations performed by \cite{Chevallard_2013}, as well as with other global galaxy observations \citep[e.g,][]{Salmon_2016,Leja_2017,Narayanan_2018}. Overall, we feel confident in asserting that the observed $A_{V_{eff}}$-$R_{V_{eff}}$ relation is indeed driven by the observation angle, with factors such as dust density and galaxy type contributing mainly to an increase in the trend dispersion. 

\subsubsection{Local observations}
\label{sec:analysis_local}
\begin{figure*}
	\centering
	\includegraphics[width=\textwidth]{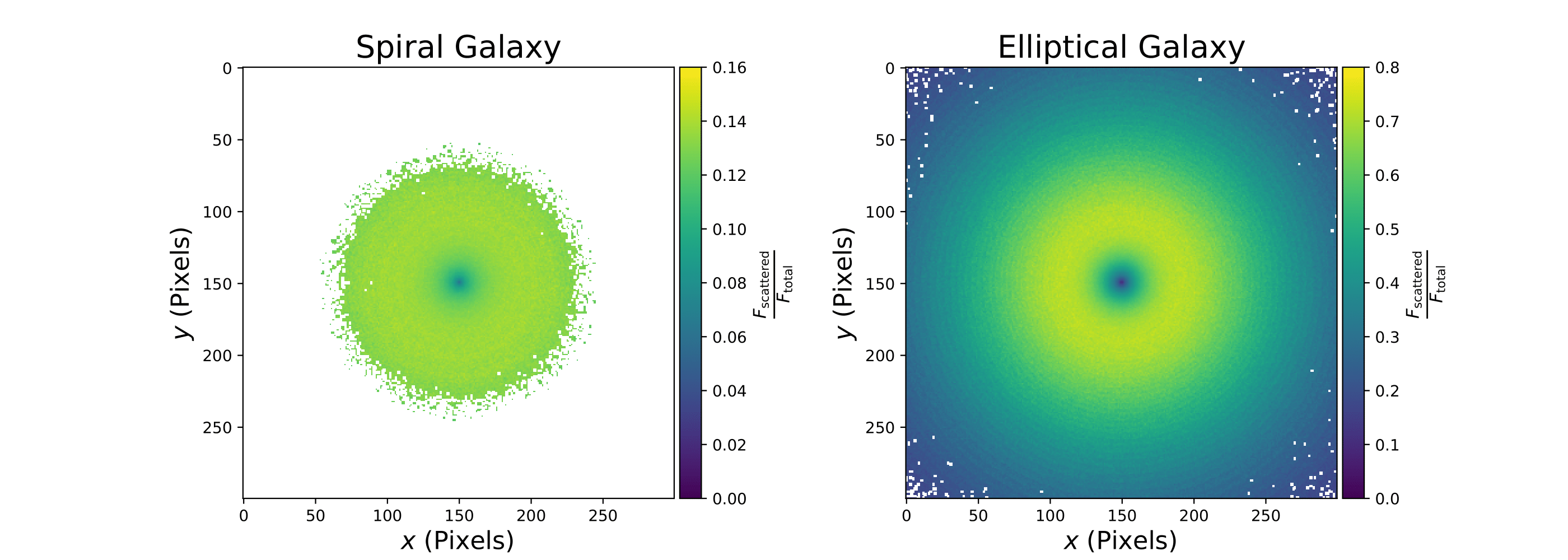}
	\caption{Spatial maps of the ratio between scattered and total observed flux for a wavelength of $\lambda=0.551\si{\micro\meter}$, for a simulated spiral (left) and elliptical face-on galaxy (right). A pixel size  corresponding to a $25$ pc resolution was used. The galaxy dust density has been normalized so that the total optical depth along the R axis is $\tau_{V_R}=5$. Only pixels with $A_{V_{eff}}>0$ and $R_{V_{eff}}>0$ are shown.
 }
\label{fig:scatter_IFU}
\end{figure*}

We can further explore the spiral and elliptical models by performing a full-galaxy field (pixel by pixel) observation of both types of galaxies. The first question we want to answer is whether a pixel size corresponding to a $25$ pc resolution is enough to achieve an extinction regime, similar to what was observed for the point source in Fig. \ref{fig:rv_aperture}. To this effect, in Fig. \ref{fig:scatter_IFU} we show maps for the scattered-to-total flux ratios for a wavelength of $\lambda=0.551\si{\micro\meter}$, which was chosen as an illustrative value. These maps were obtained from face-on simulations of the two galaxy types discussed above. The cloud dust density has been normalized so that the total optical depth along the R axis is $\tau_{V_R}=5$.

\begin{figure*}
	\centering
	\includegraphics[width=\textwidth]{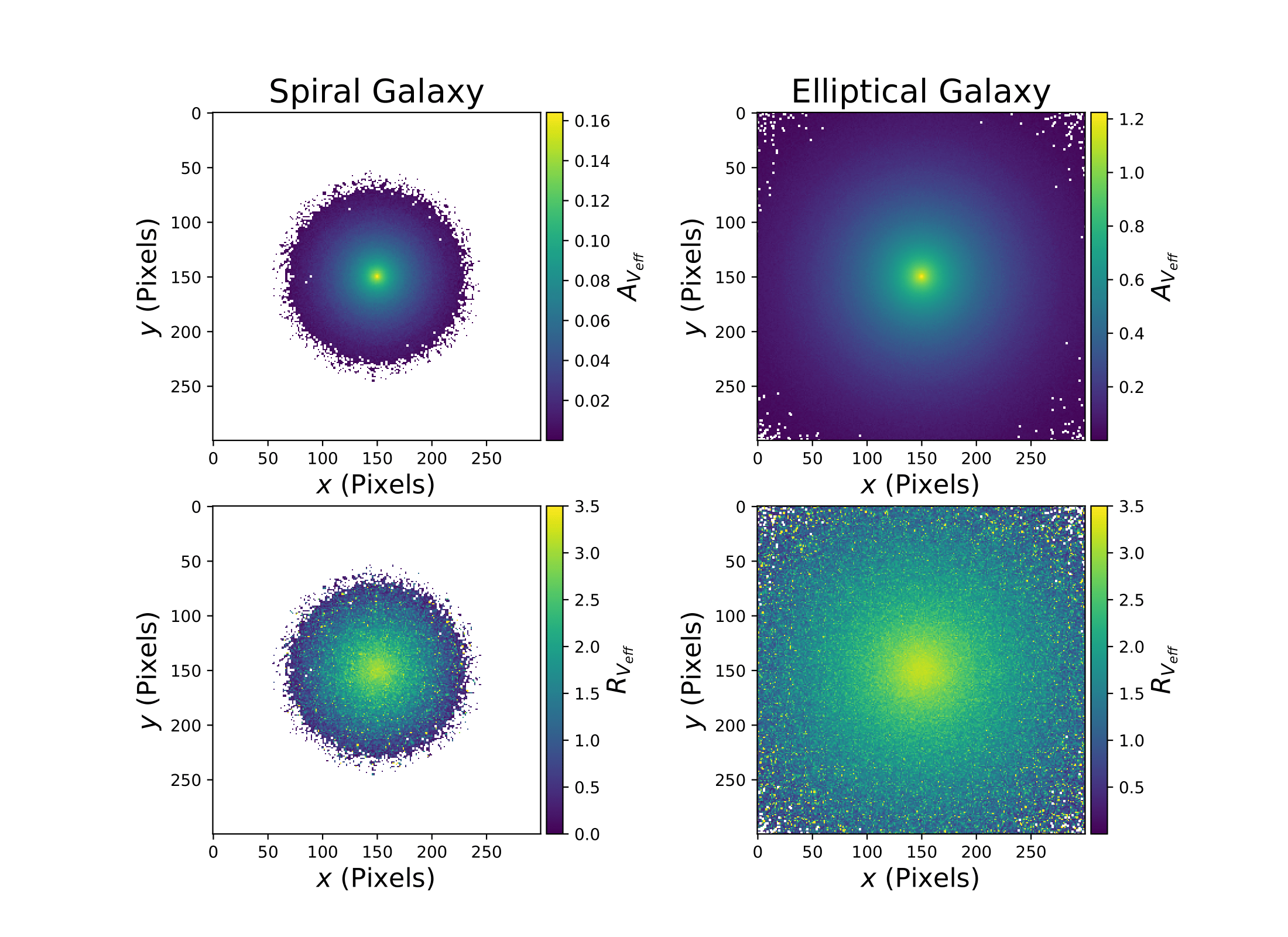}
	\caption{Spatial maps of the values of $A_{V_{eff}}$ and $R_{V_{eff}}$, for a simulated spiral (left) and elliptical  face-on galaxy (right). A pixel size corresponding to a $25$ pc resolution was used. The galaxy dust density has been normalized so that the total optical depth along the R axis is $\tau_{V_R}=5$. Only pixels with $A_{V_{eff}}>0$ and $R_{V_{eff}}>0$ are shown.}
\label{fig:av_rv_cube_IFU}
\end{figure*}

Contrary to what we found for a point source, observing with a $25$ pc resolution appears to be insufficient to meaningfully mitigate the scattered flux component across the entire galaxy, which now comes from integrated light. Even so, it appears that, for regions such as the centers of the galaxies, where the dust density is high, the scattered light is indeed negligible. In fact, looking at Fig. \ref{fig:av_rv_cube_IFU}, where maps of $A_{V_{eff}}$ and $R_{V_{eff}}$ for these two galaxies are shown, we see that, for these central regions, we recover a value of $R_{V_{eff}}$ close to $R_{V_{Zubko}}$. As we move away from the center of the galaxy, we find a decrease in the values of $R_{V_{eff}}$. This is consistent with the gradient recovered by \cite{Hutton_2015}, who find that the observed $R_{V_{eff}}$ values decrease with the galactocentric distance, when using both ultraviolet and optical photometry. While the authors interpret this as evidence for a radial variation in intrinsic dust properties in galaxies, Fig. \ref{fig:av_rv_cube_IFU} shows that the same behavior can be recovered with a uniform dust type as a consequence of different levels of scattering and dust-to-stars distributions. There are also some outlier pixels with uncharacteristically high $R_{V_{eff}}$ values, but these arise mainly due to stochastic variations in the simulation and are particularly relevant for thin dust layers, such as the one present in the spiral edge-on view. 

\begin{figure*}
	\centering
	\includegraphics[width=\textwidth]{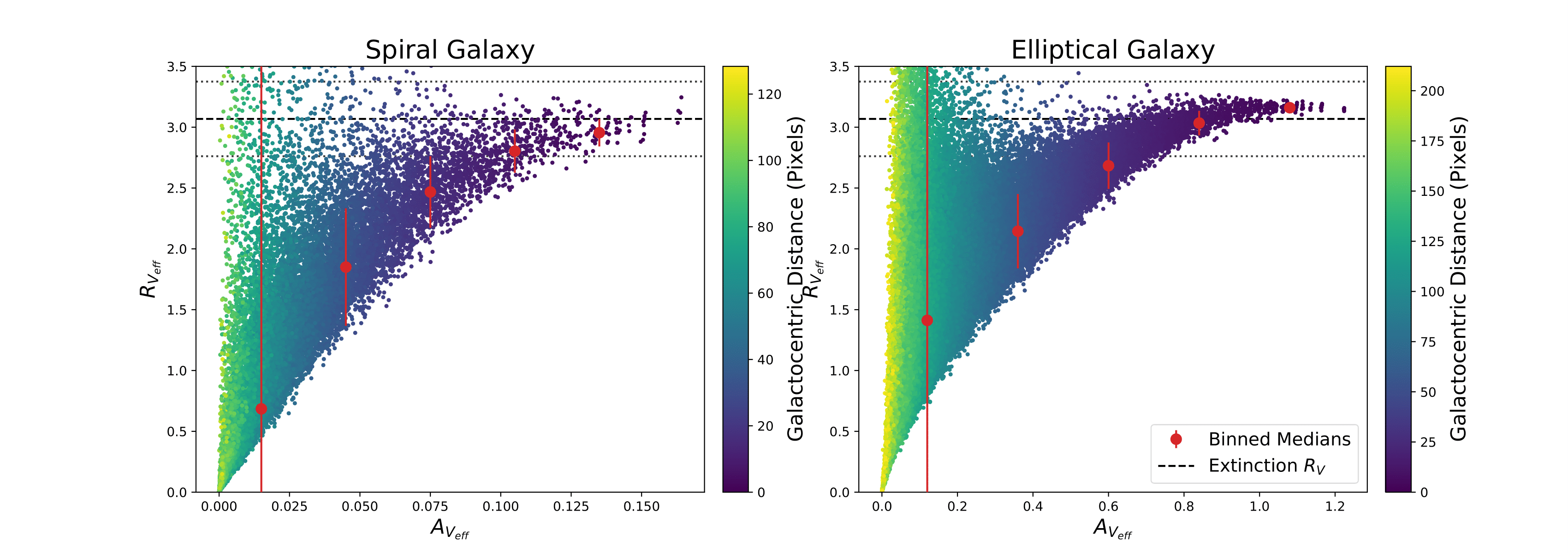}
	\caption{2D maps of the values for $A_{V_{eff}}$ and $R_{V_{eff}}$ for a face-on simulated spiral (left) and elliptical galaxy (right). The galaxy dust density has been normalized so that the total optical depth along the R axis is $\tau_{V_R}=5$. Each point corresponds to a single pixel with a resolution of $25$ pc, color-coded according to their galactocentric distance. Binned medians are shown in red, with error-bars defined by the bin standard deviation. The dashed black line reflects the true $R_V$ of the medium, with the dotted black lines defining a 10\% deviation from this value. Only pixels with $A_{V_{eff}}>0$ and $R_{V_{eff}}>0$ are shown.}
\label{fig:av_vs_rv_IFU}
\end{figure*}

The relation between the two dust parameters becomes even more apparent if we look at the $A_{V_{eff}}-R_{V_{eff}}$ trends for both face-on galaxy models, plotted in Fig. \ref{fig:av_vs_rv_IFU}. As was stated above, in addition to the light scattered into the line of sight, which tends to be reddened and contributes to drive $R_{V_{eff}}$ down, one must also consider the existence of multiple sources at different depths in the same optical path, which, when taken together, contribute to an increase in $R_{V_{eff}}$. For higher values of $A_{V_{eff}}$, and thus higher dust densities, the amount of light that can escape from the innermost parts of the galaxy, be it scattered or otherwise, drastically goes down. As such, the observed flux becomes largely dominated by the direct flux of stars located in the outermost layers of the galaxy, which mitigates both of the above-mentioned effects.

\par
For this reason, we can recover a value of $R_{V_{eff}}$ within $10\%$ of $R_{V_{Zubko}}$ when observing areas with large optical depth near the center of the galaxies. This recovery is also aided by the high central source density, which also contributes to make the scattered component subdominant in this particular region. We end up slightly overestimating the true extinction value mainly because it is impossible to completely eliminate the contributions from sources deeper in the galaxy, which slightly increased the values of $R_{V_{eff}}$, as is seen for the stellar population modeled here. 
\par
\begin{figure*}
	\centering
	\includegraphics[width=\textwidth]{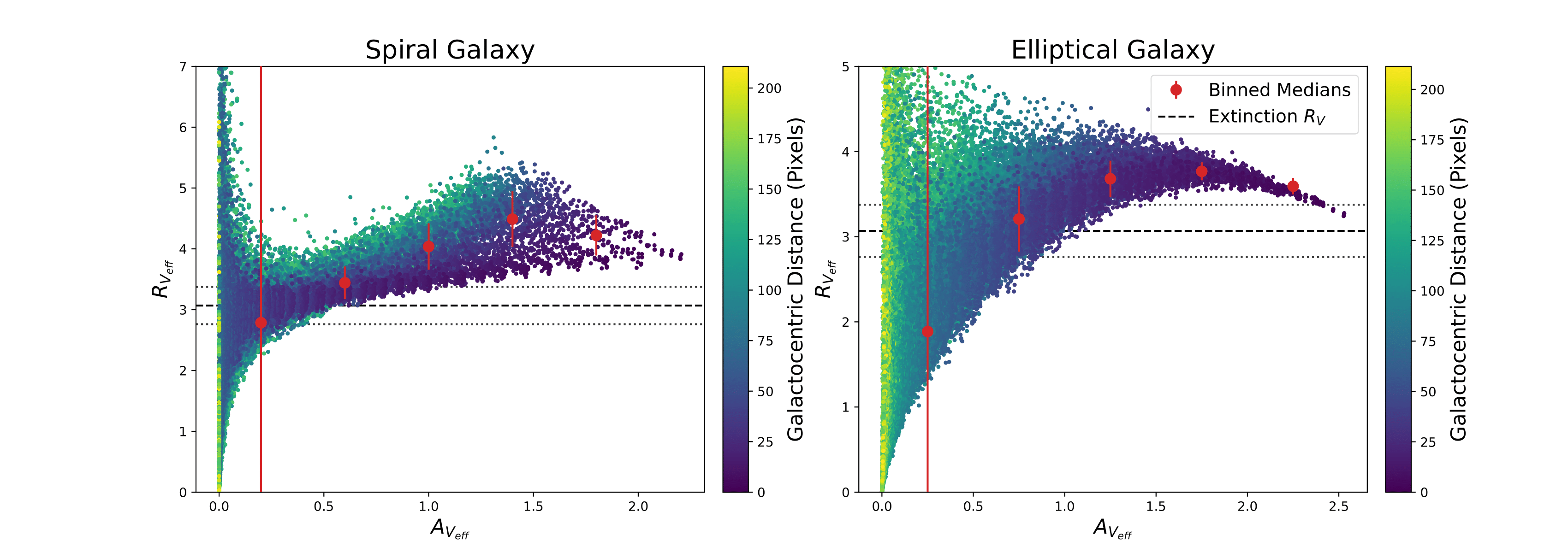}
	\caption{Similar to Fig. \ref{fig:av_vs_rv_IFU} for simulated edge-on spiral (left) and elliptical galaxy (right). For the spiral galaxy, the true extinction $R_V$ of the medium appears to be better traced by the low $A_{V_{eff}}$ regions, instead of the highly attenuated ones.}
\label{fig:av_vs_rv_IFU_edge}
\end{figure*}

Looking at the edge-on case for the same two models, plotted in Fig. \ref{fig:av_vs_rv_IFU_edge}, we see a very similar picture, especially for the elliptical galaxy. The $R_{V_{eff}}$ values are overall higher, as expected from an increase in dust cloud depth and attested by Fig. \ref{fig:av_vs_rv_norms}, but converge toward $R_{V_{Zubko}}$ for the highest $A_{V_{eff}}$ regions. The spiral galaxy case presents a few more difficulties, as the $R_{V_{eff}}$ trend never converges to within $10\%$ of $R_{V_{Zubko}}$. The most likely explanation for this is that light from the deepest layers of the galaxy is not being sufficiently attenuated for the outermost stars to dominate the emission, leading to $R_{V_{eff}}$ values much higher than $R_{V_{Zubko}}$, even for regions with higher $A_{V_{eff}}$. In fact, it appears that, despite the high level of dispersion, the low $A_{V_{eff}}$ regions are much better at tracing $R_{V_{Zubko}}$, possibly because these are also the regions with the fewest stars. Overall, it appears that, to ensure that the $R_{V_{eff}}$ values recovered for a galaxy are indeed reflective of $R_{V_{Zubko}}$, we must be sure that both the scattered flux component and the direct component coming from deeper regions of the galaxy are negligible. 

\par
\begin{figure}
	\centering
	\includegraphics[width=0.5\textwidth]{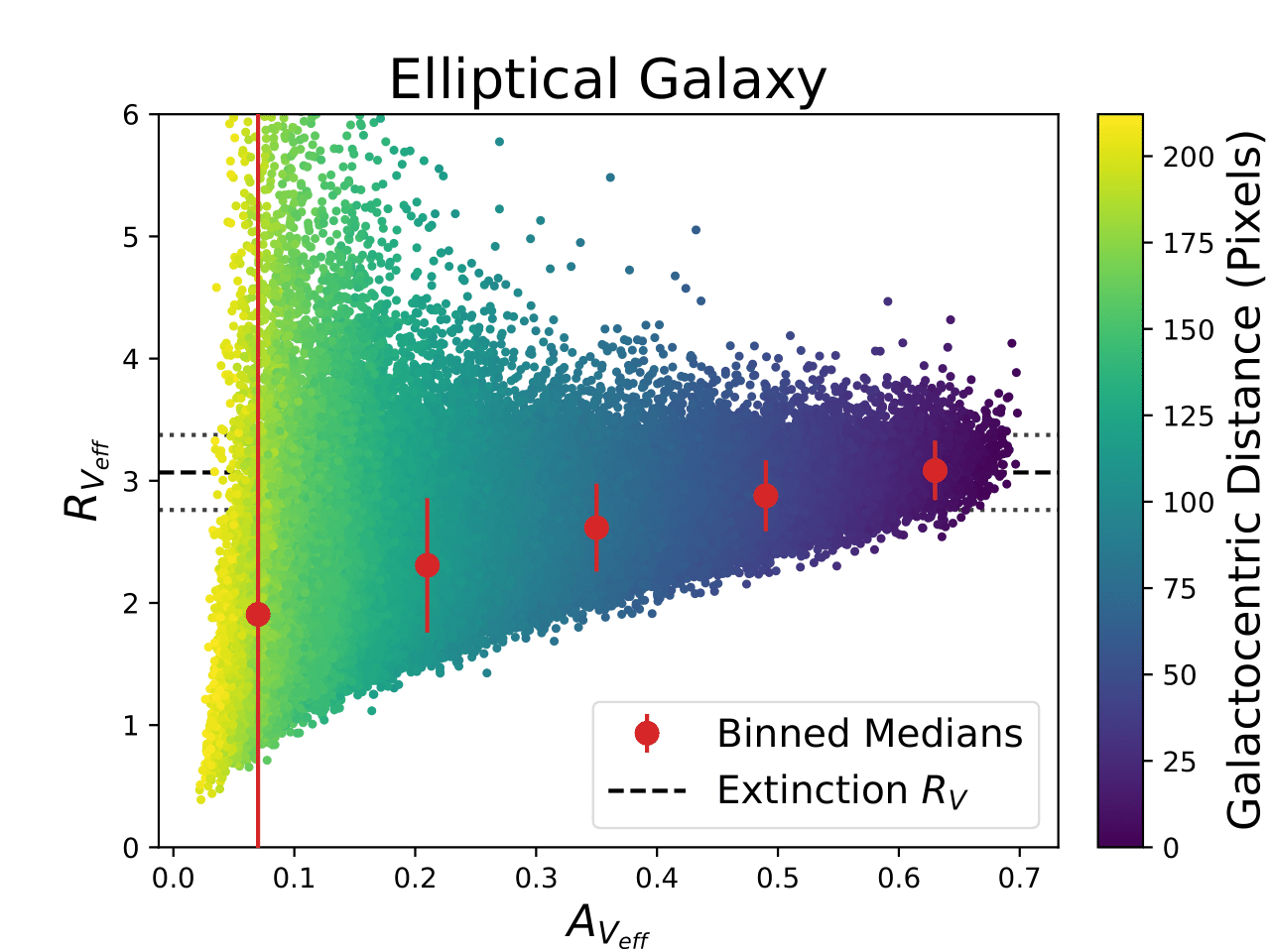}
	\caption{Similar to Fig. \ref{fig:av_vs_rv_IFU} for simulated face-on elliptical galaxy with a source distribution given by a Sérsic profile with $n=1$.}

\label{fig:av_vs_rv_IFU_sersic_1}
\end{figure}
\begin{figure}
	\centering
	\includegraphics[width=0.5\textwidth]{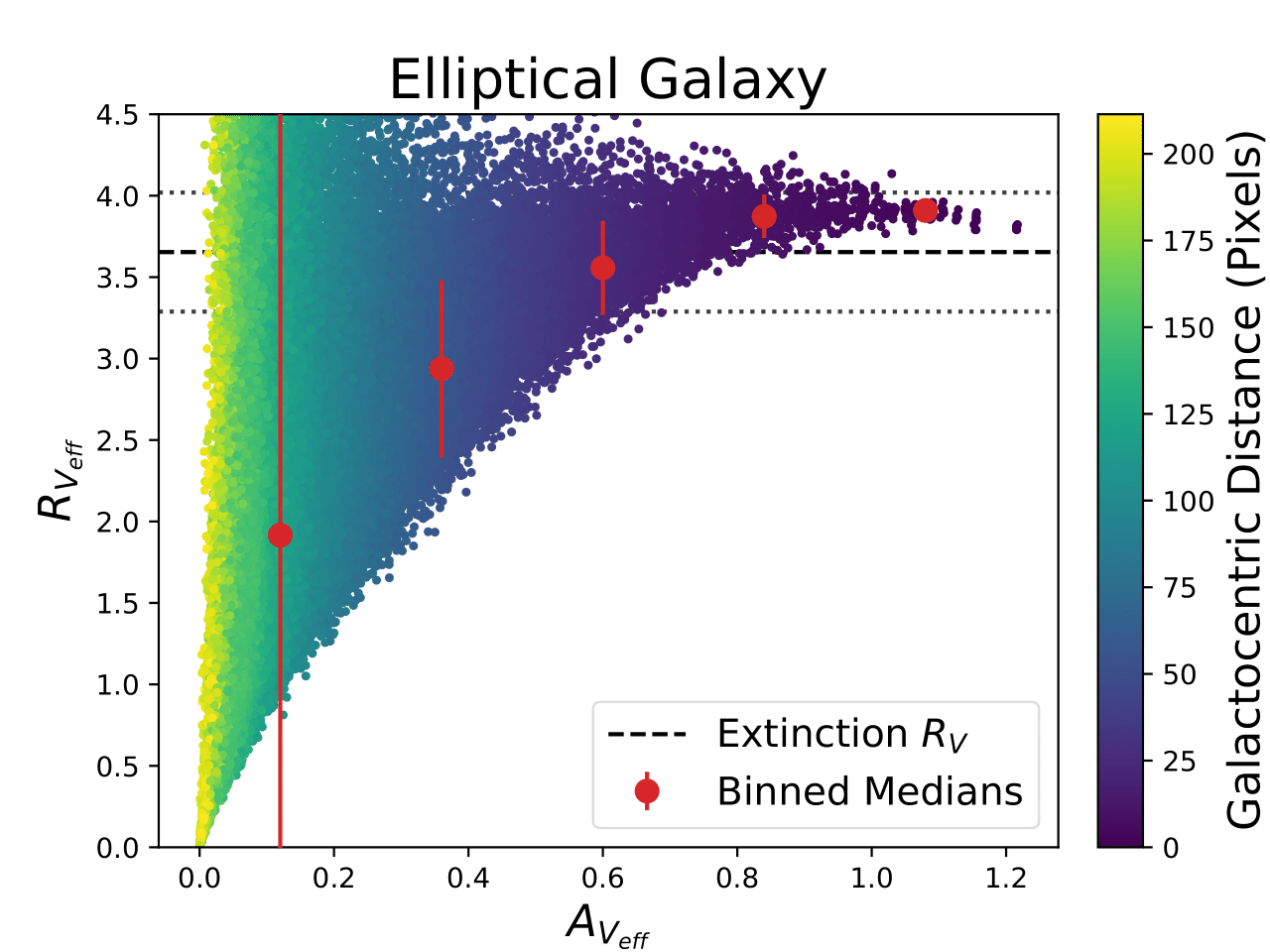}
	\caption{Similar to Fig. \ref{fig:av_vs_rv_IFU} for a simulated face-on elliptical galaxy with a MRN dust medium.}
\label{fig:av_vs_rv_IFU_MRN}
\end{figure}

From the individual examples discussed above, however, it is not completely clear whether this apparent recovery of $R_{V_{Zubko}}$ for high $A_{V_{eff}}$ regions is to be generally expected or whether it is an artifact of the interplay between these specific dust and source configurations. Thus, we further investigate this effect by examining two slightly modified examples of the elliptical face-on simulation: i) an elliptical galaxy with a lower central source density, which can be achieved using a Sérsic index $n=1$; ii) an elliptical galaxy with a MRN dust mix, instead of a Zubko dust mix, which has an intrinsic extinction $R_{V_{MRN}}=3.65$. The results for the $A_{V_{eff}}-R_{V_{eff}}$ trends, are displayed in Figs. \ref{fig:av_vs_rv_IFU_sersic_1} and \ref{fig:av_vs_rv_IFU_MRN}, respectively.

\par

Despite a much higher level of scatter, Fig. \ref{fig:av_vs_rv_IFU_sersic_1} shows that, for higher values of $A_{V_{eff}}$, $R_{V_{eff}}$ continues to approach $R_{V_{Zubko}}$ for the lower central density source distribution. Likewise, using a MRN dust mix substantially changes the value toward which the $R_{V_{eff}}$ trend converges, which is once again very similar to the extinction curve slope value for this particular mix, $R_{V_{MRN}}$.
\par
From these examples we conclude that, to recover a median $R_{V_{eff}}$ within a $10\%$ deviation of the extinction $R_V$ in a face-on galaxy, one must focus on the more attenuated areas of a galaxy. The $A_{V_{eff}}$ value for which this threshold is met can range from $40\%$ of the maximum galaxy $A_{V_{eff}}$, in the case of the elliptical galaxy with MRN dust, to $65\%$, in the case of the spiral galaxy. We highlight that these values were derived from a small number of example galaxies and should be taken as merely indicative and that they can vary with observation angle.
\par
One other thing that should be considered is that a single galaxy might host different dust types. In this case, and assuming that the different types are confined to different spatial regions, it is not correct to take the globally most attenuated region as a proxy for the extinction behavior of the whole galaxy. Rather, the most attenuated portions in each region must be considered, with the caveat that it is possible that these will not be dense enough to get within a $10\%$ deviation of the extinction $R_V$. Additionally, we must be certain that the interplay between sources and dust in the regions in question is such that the scattered flux is negligible.
\par
We can examine how well these results line up with observations by examining some preliminary dust attenuation values obtained for the All-weather MUse Supernova Integral Field Nearby Galaxies \citep[AMUSING;][]{Galbany_2016} data for the NGC 1448 galaxy. This integral field spectroscopy (IFS) data was matched to GALEX \citep{Martin_2005}, HST, 2MASS \citep{Skrutskie_2006} and WISE \citep{Wright_2010} photometry using the piXedfit Python package \citep{Abdurro'uf_2021}. The Prospector Python package \citep{Johnson_2021} was used to fit each pixel in a subsection of NGC 1448, with the attenuation properties being expressed in terms of the effective optical depth, $\tau_{V_{eff}}$, and the dust index, $n$, which controls the slope of the attenuation curve in \cite{Kriek_2013}, such that, on a rough approximation, we have $R_{V_{eff}}\sim\left(\frac{\lambda_V}{\lambda_B}\right)^n$. As was mentioned in Section \ref{sec:aperture_galaxy}, there appears to exist a bias in Prospector that results in the overestimation of $\tau_{V_{eff}}$. This does not pose a problem to our analysis, as we are primarily interested in the $\tau_{V_{eff}}$-$n$ trends.
\par
\begin{figure}
	\centering
	\includegraphics[width=0.5\textwidth]{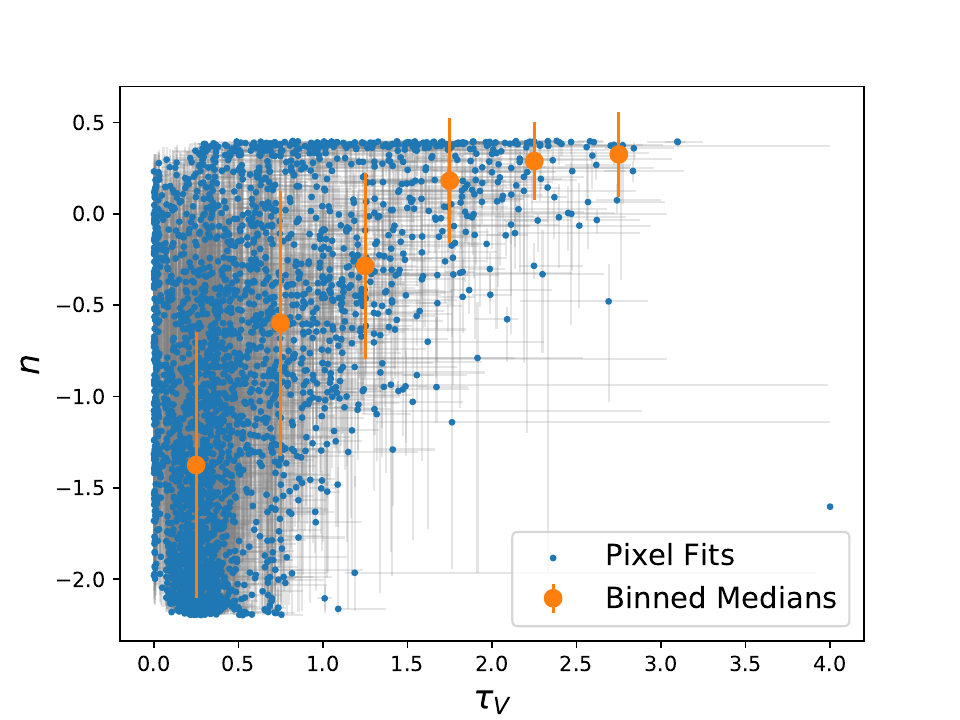}
	\caption{Dust index, $n$, as a function of the optical depth $\tau_{V_{eff}}$ for fits of a subsection of NGC 1448, using AMUSING IFS data matched with GALEX, HST, 2MASS and WISE photometry. Values for individual pixels are shown in blue, while binned median values are shown in orange.}
\label{fig:muse_comp}
\end{figure}

The  results for the fitted $\tau_{V_{eff}}$-$n$ trends are displayed in Fig. \ref{fig:muse_comp}. While these fits are still preliminary, it appears that the same trend observed in Fig. \ref{fig:av_vs_rv_IFU}, \ref{fig:av_vs_rv_IFU_sersic_1} and \ref{fig:av_vs_rv_IFU_MRN} is present in the AMUSING data, even if expressed through different parameters. It is not totally clear whether the $n$ values are converging toward a value representative of the extinction properties of the medium, or if they are simply clustering near the edge of the prior distribution. Furthermore, our simulations and fits match the results obtained by \cite{Decleir_2019}, who find a similar increase in the attenuation curve slope with $A_{V_{eff}}$ for resolved observations of NGC 628. These results give some confidence to the applicability of the inferred simulated trends to real galaxies.

\section{Discussion}
\label{sec:analysis}
\subsection{Dust attenuation and stellar mass}
\label{sec:dis_mass}
Given the nature of the input of the SKIRT simulations, where both the source and dust medium systems require their normalizations to be specified as inputs, a direct probing of the relation between galaxy mass and dust cannot be achieved, as the relation between the two is always determined by the user. However, some important inferences can be drawn from the comparison of our simulations with literature results.
\par
Among others, \cite{Garn_2010}, \cite{Zahid_2013}, \cite{Salim_2018}, \cite{Galbany_2022} and \cite{Duarte_2022} show that, for star-forming galaxies, there is an increase in $A_{V_{eff}}$ with stellar mass. It is fair to assume that this is related to an increase in the dust density in high-mass galaxies. \cite{Salim_2018} and \cite{Duarte_2022} also show that high-mass star-forming galaxies exhibit larger values of $R_{V_{eff}}$. From Fig. \ref{fig:av_vs_rv_norms} it is clear that an increase in the dust density in a galaxy leads directly to an increase in the median values of both $A_{V_{eff}}$ and $R_{V_{eff}}$. Assuming that there is indeed a link between the mass of a galaxy and its dust density, our simulations are able to fully explain the observed increase in $R_{V_{eff}}$ with stellar mass. Crucially, they do so without the need to consider the differing dust types or source SEDs.

\par
Another important result discussed in \cite{Zahid_2013}, \cite{Salim_2018} and \cite{Duarte_2022} is that elliptical galaxies are, in general, expected to have much lower values of $A_{V_{eff}}$ and $R_{V_{eff}}$ than high-mass star-forming galaxies. It is this fact that makes it possible for the entire high-mass population to actually have a lower median $R_{V_{eff}}$ than the one observed for the low-mass galaxies. However, this is something our simulations cannot reproduce. If one assumes that the median $A_{V_{eff}}$ observed for a population of elliptical galaxies should be roughly the same as the one observed for a population of low-mass star-forming galaxies, which is corroborated by \cite{Zahid_2013}, \cite{Salim_2018} and \cite{Duarte_2022}, we can use this assumption to properly normalize the dust contents of both galaxy types. However, under these conditions, we invariably find that the ellipticals have median $R_{V_{eff}}$ values higher than even those of the high-mass star-forming galaxies. There are two likely explanations. The first is that the simplistic geometries assumed in this work are not able to properly model these two types of galaxies in relation to one another. The second is that there is some fundamental difference between the two galaxy types that goes beyond geometry, such as a different dust composition with different optical properties, which makes it so that ellipticals have intrinsically lower $R_V$ values. This would support the claim that multiple dust properties can be present in SN Ia cosmological analyses, as has been argued by, among others, \cite{Brout_2021}, \cite{gaitan_2021}, and \cite{wiseman_2022}.

\par
As is suggested in Figs. \ref{fig:av_vs_rv_IFU}, \ref{fig:av_vs_rv_IFU_sersic_1}, and \ref{fig:av_vs_rv_IFU_MRN} and tentatively confirmed by Fig. \ref{fig:muse_comp}, the analysis of full-galaxy field data for the regions of a face-on galaxy with higher $A_{V_{eff}}$ might offer an observational hint at the true extinction curve slope $R_V$ of a given dust sample. While the robustness of this methodology in determining accurate $R_V$ values for real galactic dust mediums cannot be ascertained from our simulations, observational results are promising, as they show a similar increase in $R_{V_{eff}}$ with $A_{V_{eff}}$ as the one recovered for the simulations. Thus, the analysis of the local dust properties in highly attenuated regions of face-on galaxies might be a very apt method to distinguish between different dust types, perhaps even within the same galaxy. As such, it might be able to provide a reliable way to relate galaxy attenuation data with the extinction of their hosted point sources, such as SNe. This would be invaluable in an effort to better constrain SN Ia standardization and the respective color-luminosity correction.

\subsection{Source SED impact}
\label{sec:SED_impact}

As was mentioned above, when it comes to dust, the results presented in this paper are largely independent of the SED of the simulated sources, given that our simulations output precise fluxes for the emitted and observed fluxes. Because of this, we can avoid many of the uncertainties and degeneracies that are inherent in observations and SED fitting. As an example, as was stated in Section \ref{sec:analysis_point_source}, we recover the exact same values of $A_{V_{eff}}$ and $R_{V_{eff}}$ when using light source SEDs with either $t_{age}=10$ Myr or $t_{age}=5$ Gyr. This happens because the general scattering and attenuation effects found in this study are experienced by all sources of light, irrespective of their SED. However, there are a few points to be highlighted.
\par
One of the reasons why our results can be said to be SED independent is that, for any given model simulation, a single SED is used. This means that any effect of the source is factored out when the effective dust properties are computed. We do not, however, expect real galaxies to exhibit a SED that is uniform across all sources, i.e., we do not expect stellar populations of the same age everywhere. In this scenario, both $A_{V_{eff}}$ and $R_{V_{eff}}$ are also expected to be affected by the interplay of the various SEDs, particularly if sources of very different ages are not uniformly distributed within the galaxy and are subject to different levels of attenuation. We examine two specific examples to further illustrate this point.

\par
We first look at a simulated spiral galaxy similar to the one described in Section \ref{sec:spirals}, but with disk age $t_{disk}=100$ Myr and bulge age $t_{bulge}=5$ Gyr. Although not totally realistic, this simulation helps illustrate the impact of multiple SEDs differently distributed on the recovered effective dust properties. The 2D face-on and edge-on views for this galaxy are similar to those obtained for a spiral galaxy with identical morphology and a single SED (see Figs. \ref{fig:av_vs_rv_IFU} and \ref{fig:av_vs_rv_IFU_edge}). The main exception is that, in the edge-on view of the two-population simulation, $R_{V_{eff}}$ diverges much more from $R_{V_{Zubko}}$ as $A_{V_{eff}}$ increases.
\par
\begin{figure}
	\centering
	\includegraphics[width=0.5\textwidth]{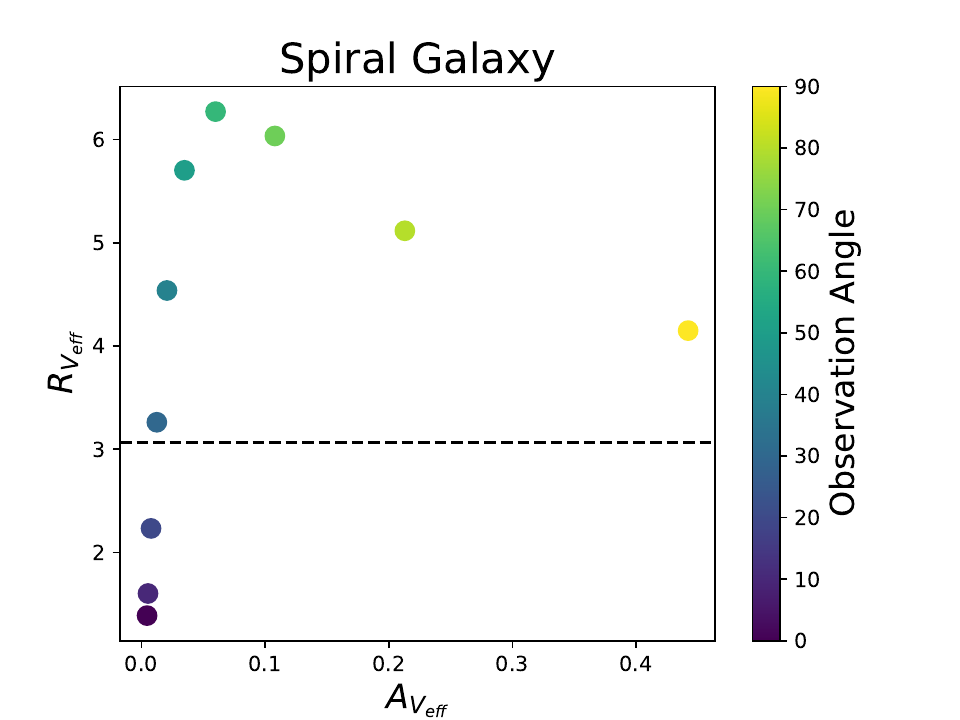}
	\caption{Values of $R_{V_{eff}}$ as a function of $A_{V_{eff}}$ for a simulated spiral galaxy with disk age $t_{disk}=100$ Myr and bulge age $t_{bulge}=5$ Gyr. Different colors denote different observation angles, from $0^\circ$ (face-on) to $90^\circ$ (edge-on). An aperture radius of $20$ kpc was used. The galaxy dust density has been normalized so that the total optical depth along the R axis is $\tau_{V_R}=5$. The dashed black line reflects the true extinction $R_{V_{Zubko}}$.}
    
\label{fig:old_bulge}
\end{figure}

Regarding the variation in the global dust properties across various observing angles, plotted in Fig. \ref{fig:old_bulge}, we find that it substantially differs from the results displayed in Fig. \ref{fig:av_vs_rv_norm5}. The presence of different star populations contributes to further complicate the geometric effects on the observed light, making it even harder to establish a link between extinction and attenuation. The fact that observed galaxy dust data are closer to the single SED simulation might indicate that the differences in age distributions between populations of the same galaxy are not as large as assumed in this extreme example.

\par
\begin{figure}
	\centering
	\includegraphics[width=0.5\textwidth]{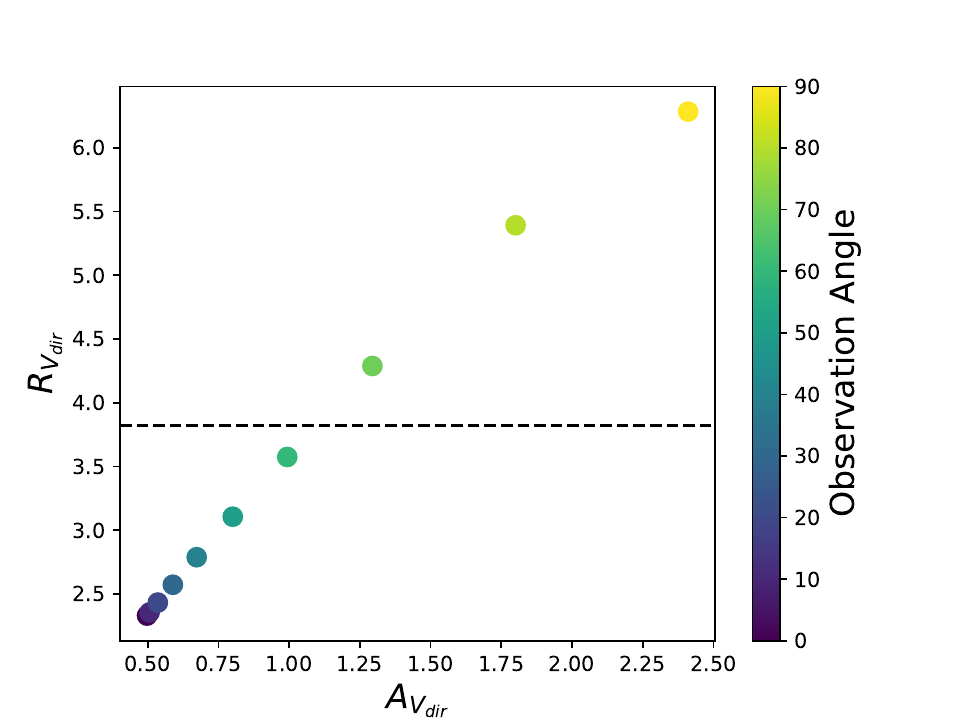}
	\caption{Values of $R_{V_{eff}}$ as a function of $A_{V_{eff}}$ for a simulation based on the M51 galaxy from DustPedia. Different colors denote different observation angles, from $0^\circ$ (face-on) to $90^\circ$ (edge-on). An aperture radius of $20$ kpc was used. The galaxy dust density has been normalized so that the total optical depth along the R axis is $\tau_{V_R}=5$. The dashed black line reflects the true extinction $R_{V_{THEMIS}}$.}
    
\label{fig:M51}
\end{figure}

For a more realistic example, we examined a simulation based on the M51 galaxy model from the DustPedia\footnote{\url{http://dustpedia.astro.noa.gr/}} archive and uses observations of M51 to define both the stellar and dust morphologies. Crucially, the model contains both old and young stars, with a much more realistic distribution than the one used for the previous example. The dust medium is characterized by a THEMIS dust mix \citep{Jones_2017}, with dust mass $M_{dust}=10^8M_{\odot}$. The global dust properties across various observing angles for the simulated M51 galaxy are plotted in Fig. \ref{fig:M51}. The results for this more realistic two-population simulation are much more in line with the ones recovered for both the single-population simulations and observations. This means that, for real galaxies, while the mixing of different SEDs might influence the individual values of $R_{V_{eff}}$, it will most likely not have a big impact on the overall shape of the $A_{V_{eff}}-R_{V_{eff}}$ trend.

\section{Conclusions}
In this work we have used radiative transfer simulations to explore how SN Ia dust extinction and host galaxy dust attenuation relate to one another. More specifically, we have investigated under which conditions, if any, attenuation can be used as a proxy for extinction. Our main findings are the following:
\begin{enumerate}[(i)]
    \item SN Ia extinction (or ``point source extinction'' as used throughout this paper) and host galaxy attenuation are fundamentally different phenomena. When observing a galaxy, the light scattered into the line of sight becomes a very relevant component of the total observed flux, lowering the value of the effective $R_{V_{eff}}$ with respect to the medium's intrinsic extinction $R_V$. In addition, sources at different depths contained along the same line of sight contribute to an increase in $R_{V_{eff}}$. When taken in conjunction, these two effects mean that the $R_{V_{eff}}$ measured for any given galaxy is fundamentally different from the optical extinction $R_V$ of its particular dust medium, even if small resolutions of $25$ pc are considered. These effects are further exacerbated if distinct stellar populations with different SEDs and spatial distributions are present in the galaxy, as is likely to occur. Thus, SN Ia standardization cannot be improved by using global or local $R_{V_{eff}}$ values obtained from SED stellar population fits to host galaxy data as proxies for the extinction of individual SNe. Doing so will only contribute to further bias the results, due to the geometric effects inherent to galaxy observations.
    
    \item Global attenuation properties for the simulated galaxies are primarily driven by the observation angle. The global $A_{V_{eff}}-R_{V_{eff}}$ trends for observed galaxies, which have different morphologies, dust contents and stellar masses, can be qualitatively reproduced by a single simulated galaxy, with a single dust type, observed from different angles. This shows that the importance of geometric effects cannot be neglected in any analysis of galactic dust.
    
    \item Galaxy dust density, as parameterized by the cloud normalization in the simulation, is directly linked to $A_{V_{eff}}$ and $R_{V_{eff}}$. Increasing the dust density leads to an increase in both quantities across all observation angles. As such, if a correlation between stellar mass and dust density is assumed, the geometric effects in the observations are also able to explain the observed increase in $R_{V_{eff}}$ values with stellar mass for star-forming spiral galaxies.
    
    \item When spiral and elliptical galaxies are analyzed together, our simulations with a single dust type fail to explain some of the observed behaviors. The literature suggests that ellipticals have $A_{V_{eff}}$ values similar to low-mass spirals. However, if this is assumed in our model, the recovered $R_{V_{eff}}$ values, which are also supposed to be in line with those of low-mass spirals, are much higher than expected. This means that there is some intrinsic difference between the two galaxy types that cannot be modeled by geometrical and morphological effects, and that particularly relates to the optical properties of their specific dust types, with ellipticals having intrinsically lower $R_V$ values. In this regard, there could be multiple distinct dust populations associated with different galaxy populations that need to be corrected for accordingly in SN Ia standardization, as has previously been suggested in the literature.
    
    \item All the geometrical effects contributing to $R_{V_{eff}}$ make it difficult to extract accurate information about intrinsic dust properties, even for well-resolved observations. However, our simulations suggest that, for elliptical and face-on spiral galaxies, one can recover $R_{V_{eff}}$ values within a $10\%$ error of the intrinsic extinction $R_V$ for sufficiently resolved areas with high $A_{V_{eff}}$. For edge-on spirals, the median $R_{V_{eff}}$ for low-$A_{V_{eff}}$ regions, although with a lot of scatter, appears to be better at tracing the intrinsic extinction $R_V$. This conclusion is tentatively supported by preliminary fits at resolved location of real observed galaxies, although it is difficult to confirm without direct knowledge of the intrinsic optical properties of the dust. In any case, it is crucial to ensure that the regions chosen to proxy extinction in SNe Ia or any other point sources allow for the elimination of both the scattered and the direct flux components coming from deeper regions of the galaxy. In this way, it might ultimately be possible to relate galaxy attenuation to the extinction of their hosted SNe Ia, which would allow individual SNe dust properties to be independently corrected. This would greatly improve upon SN Ia standardization, as it would mean forgoing the often assumed universal color-luminosity correction.

\end{enumerate}

\label{sec:conclusions}

\begin{acknowledgements}
J. D., S. G. G. and A. M. acknowledge support by FCT for CENTRA through the Project No. UIDB/00099/2020. J. D. also acknowledges support by FCT under the PhD grant 2023.01333.BD, with DOI https://doi.org/10.54499/2023.01333.BD. S.G.G. acknowledges support from the ESO Scientific Visitor Programme.

This work was funded by ANID, Millennium Science Initiative, ICN12\_009.

M. S. acknowledges support by the Ministry of Science, Technological Development and Innovation of the Republic of Serbia (MSTDIRS) through contract no. 451-03-136/2025-03/200002 with the Astronomical Observatory (Belgrade).

Based in part on observations collected at the European Southern Observatory under ESO programme(s) 097.D-0408(A). 

Based in part on observations made with the NASA/ESA Hubble Space Telescope, obtained at the Space Telescope Science Institute, which is operated by the Association of Universities for Research in Astronomy, Inc., under NASA contract NAS5-26555, associated with programs \#17070 and \#17191.

This publication makes use of data products from the Two Micron All Sky Survey, which is a joint project of the University of Massachusetts and the Infrared Processing and Analysis Center/California Institute of Technology, funded by the National Aeronautics and Space Administration and the National Science Foundation.

This work used public data from the GALEX Survey. GALEX is operated for NASA by the California Institute of Technology under NASA contract NAS5-98034.

 This publication makes use of data products from the Wide-field Infrared Survey Explorer, which is a joint project of the University of California, Los Angeles, and the Jet Propulsion Laboratory/California Institute of Technology, funded by the National Aeronautics and Space Administration.

This work makes use of simulations from DustPedia, a collaborative focused research project supported by the European Union under the Seventh Framework Programme (2007-2013) call (proposal no. 606847). The participating institutions are: Cardiff University, UK; National Observatory of Athens, Greece; Ghent University, Belgium; Université Paris Sud, France; National Institute for Astrophysics, Italy and CEA, France.

\end{acknowledgements}

\bibliographystyle{aa}
\bibliography{aanda.bib}
\end{document}